\documentclass[preprint,floats]{revtex4}
\usepackage{graphicx,dcolumn}
\usepackage{hyperref}
\usepackage{amssymb}
\begin{document}

\title{A multiphase equation of state for carbon addressing high pressures and temperatures}

\author{Lorin X. Benedict$^{1}$, Kevin P. Driver$^{2}$, Sebastien Hamel$^{1}$, Burkhard Militzer$^{2,3}$, Tingting Qi$^{1}$, Alfredo A. Correa$^{1}$ and Eric Schwegler$^{1}$}

\affiliation{$^{1}$Condensed Matter and Materials Division, Physical and Life Sciences Directorate, Lawrence Livermore National Laboratory, Livermore, CA 94550, USA}
\affiliation{$^{2}$Department of Earth and Planetary Science, and $^{3}$Department of Astronomy, University of California at Berkeley, Berkeley, CA 94720}

\date{\today}

\begin{abstract}
We present a 5-phase equation of state (EOS) for elemental carbon. The phases considered are: diamond, BC8, simple-cubic, simple-hexagonal, and the liquid/plasma state. The solid phase free energies are constrained by density functional theory (DFT) calculations. Vibrational contributions to the free energy of each solid phase are treated within the quasiharmonic framework. The liquid free energy model is constrained by fitting to a combination of DFT molecular dynamics performed over the range 10,000 K $< T <$ 100,000 K, and path integral quantum Monte Carlo calculations for $T >$ 100,000 K (both for $\rho$ between 3 and 12 g/cc, with select higher-$\rho$ DFT calculations as well). The liquid free energy model includes an atom-in-jellium approach to account for  the effects of ionization due to temperature and pressure in the plasma state, and an ion-thermal model which includes the approach to the ideal gas limit. The precise manner in which the ideal gas limit is reached is greatly constrained by both the highest temperature DFT data and the path integral data, forcing us to discard an ion-thermal model we had used previously in favor of a new one. Predictions are made for the principal Hugoniot and the room-temperature isotherm, and comparisons are made to recent experimental results. 
\end{abstract}

\maketitle

\section{Introduction}
The high pressure and temperature equation of state (EOS) and phase diagram of carbon have received considerable attention of late. Recent laser-shock and ramp-compression studies \cite{Bradley,Nagao,Hicks,Brygoo,McWilliams,Eggert,Smith}, together with shock measurements performed with magnetically-driven flyer plates \cite{Z}, have produced data in the range from $P= 0 - 50$ Mbar. In addition, theoretical work on the EOS and phase diagram in this same range have yielded predictions which are largely (though not completely) in accord with these experimental data \cite{Correa,Z}. These studies were conducted with density functional theory (DFT) molecular dynamics (MD). Much of this recent focus on the carbon EOS, specifically in states of compression reached when starting in the diamond phase, has arisen from the interest of using high density carbon as an ablator material for capsules designed to achieve fusion at the National Ignition Facility \cite{pointdesign}. While these experimental and theoretical studies have been useful in constraining the EOS of carbon for this and related applications, it is crucial to note that the states reached by the ablator in inertial confinement fusion (ICF) are expected to include temperatures in excess of tens of eV \cite{pointdesign}. Such conditions have not yet been sufficiently characterized. 

At such high temperatures, DFT-MD is extremely challenging to perform, due to the large number of high-lying, partially-occupied single-electron states that must be included for an accurate rendering of the electronic thermal excitations. Indeed, an orbital-free variant of DFT-MD has been developed and used to handle such high temperature applications \cite{OFDFT}, but as yet, a satisfactory treatment of atomic shell structure in this approach is lacking. The approach as implemented thus far \cite{OFDFT} is (somewhat) reminiscent of a Thomas-Fermi treatment for the electrons \cite{Hora}. Another altogether different approach to describing the electronic structure of the high-$T$ plasma state is path integral Monte Carlo (PIMC) \cite{Pollock1984}. In this approach, unlike in typical implementations of DFT, there is no mean-field assumption made for the many-electron problem, and the imaginary time treatment makes high-$T$ simulations more efficient to perform than low-$T$ simulations. Although assumptions regarding the nodal surface of the many electron density matrix necessarily introduce approximations for the treatment of atomic shells, it is very encouraging that recent work on the C plasma has demonstrated that EOS predictions can be made with this method which smoothly interpolate between DFT-MD results at the lower temperatures and the high-$T$ Debye-H\"uckel limit \cite{Driver2012}.

The ultra-high pressure regime near $T= 0$ has been studied recently as well, using ab initio electronic structure techniques of the DFT variety together with a random search method to find the thermodynamically stable crystalline phases at pressures between 10 - 1000 Mbar \cite{MC}. This work extended earlier quantitative studies of the C phase diagram that treated the diamond and BC8 phases up to $P \sim 20$ Mbar \cite{PNAS}. The random search established the sequence of stable phases to be diamond $\to$ BC8 $\to$ (slight modification of) simple cubic (sc) $\to$ simple hexagonal (sh) $\to$ face-centered cubic (fcc) \cite{MC}. Though these authors did not attempt to predict the details of the C melt curve in these extreme conditions, their analysis using a quasiharmonic description of the phonons for these solid phases suggested the existence of a BC8-sc-sh triple point, as seen in Fig. 3 of their manuscript \cite{MC}.

A few years ago, Correa et al. used the available theoretical understanding of the C phase diagram at the time \cite{PNAS} to produce a three-phase EOS model for C focusing specifically on the regime of $P= 0 - 20$ Mbar and $T= 0$ - 20,000 K \cite{Correa}. The phases considered were: diamond, BC8, and the liquid. The individual phase free energy models were fit to the results of DFT-MD calculations. Since the final multiphase EOS model was designed to be used in hydrocode simulations spanning a wider range of conditions, these researchers embedded their detailed three-phase EOS model into a more coarse-grained model \cite{QEOS} that respected the ideal gas limits at high-$T$ and low-$\rho$ and the Thomas-Fermi limit \cite{Hora} at high-$\rho$. While satisfactory in a broad sense, this older EOS model suffers from three main drawbacks: 1. The embedding of the DFT-based 3-phase model into the coarse-grained model created kinks in the thermodynamic functions, even though an attempt was made to smooth out such features by interpolation. 2. There are no solid phases beyond BC8; sc and sh phases, for instance, are not included. 3. The electronic excitations of the high-$T$ liquid are treated with an average-atom Thomas-Fermi model, so atomic shell structure is not included in sufficient detail.

In this work, we remedy these deficiencies by constructing a 5-phase EOS for C, again based on ab initio calculations (for densities between $\sim$ 1 - 25 g/cc), in which the free energy models for each phase are defined over wide ranges of $\rho$ and $T$ and exhibit sensible limiting behavior. No embedding into a simpler model \cite{QEOS} is performed. The two extra phases included are the sc and sh phases. While our phase diagram differs slightly from that of Ref.\cite{MC}, the inclusion of these extra phases allows our EOS to be in accord with ab initio predictions for pressures up to well over 100 Mbar. The electronic excitations in the liquid are treated with an average-atom model called PURGATORIO which, unlike Thomas-Fermi, includes the treatment of atomic shell structure within a Kohn-Sham DFT framework \cite{PURGATORIO}. To constrain the ion-thermal model for the liquid, we perform DFT-MD in C up to 100,000 K and PIMC caculations \cite{Driver2012} from $\sim$ 200,000 K (depending on the density) to over $10^{8}$ K. We establish that the best fit to these EOS predictions requires a model in which the decay of the ion-thermal specific heat is much faster than previously expected. Our newly-developed "cell model" \cite{Correa13}, in conjunction with the PURGATORIO electron-thermal term, provides an excellent fit throughout the range of both the DFT-MD and the PIMC data. In what follows, we describe the details of our DFT and PIMC calculations, present the models we use for the free energies of the individual phases, and compare various thermodynamic tracks through our EOS with the results of recent experiments performed on high-energy laser platforms \cite{Bradley,Eggert,Smith}.

\section{Theory and Simulation}
Our computational results fall into two categories: 1. Calculations of the EOS, by which we mean internal energy, $E$, and pressure, $P$, as functions of density, $\rho$, and temperature, $T$, and 2. Calculations of intermediate quantities we which use to build the EOS. These include cold curves ($E$ and $P$ as functions of $\rho$, for ions fixed in position \cite{caveatcold}), phonon densities of states (PDOS), and electronic excitation contributions to the free energy. The first category of quantities we extract from DFT-MD (for solid and liquid phases) and PIMC (liquid phase). The second group of quantities we extract solely from DFT calculations for the solid phases. It may seem at first as if our multitude of results overdetermines the EOS, particularly in the liquid phase where both DFT-MD and PIMC results reside. However we find it crucial to make use of this full suite of results; the approximations inherent in our DFT and PIMC calculations are altogether different, and it is important to check whether certain features of our results are artifacts of the simulation method or are robust indicators of the true EOS of C. This is especially the case for the liquid at high temperatures, where we find (with both methods) results which contradict some of our previous assumptions (see below). Fig.\ref{pd_data} shows the locations of our DFT-MD and PIMC simulation points in the $(P,T)$-plane. Also shown for reference are our predicted C phase lines, presented and discussed in detail below, as well as portions of five separate isentropes computed from the EOS model described in this work. We stress that while the DFT-MD results are concentrated entirely in the diamond and liquid phases, we use the aforementioned DFT-derived intermediate quantities (cold curves, PDOSs, etc.) to constrain the EOS model for each of the four solid phases considered here. All of our DFT-MD and PIMC data, consisting of internal energy and pressure at different densities and temperatures, are reproduced in tables contained in the Supplementary Material. In the following two subsections, we describe in detail our approaches to obtaining these data.

\begin{figure}
\includegraphics[scale=0.40]{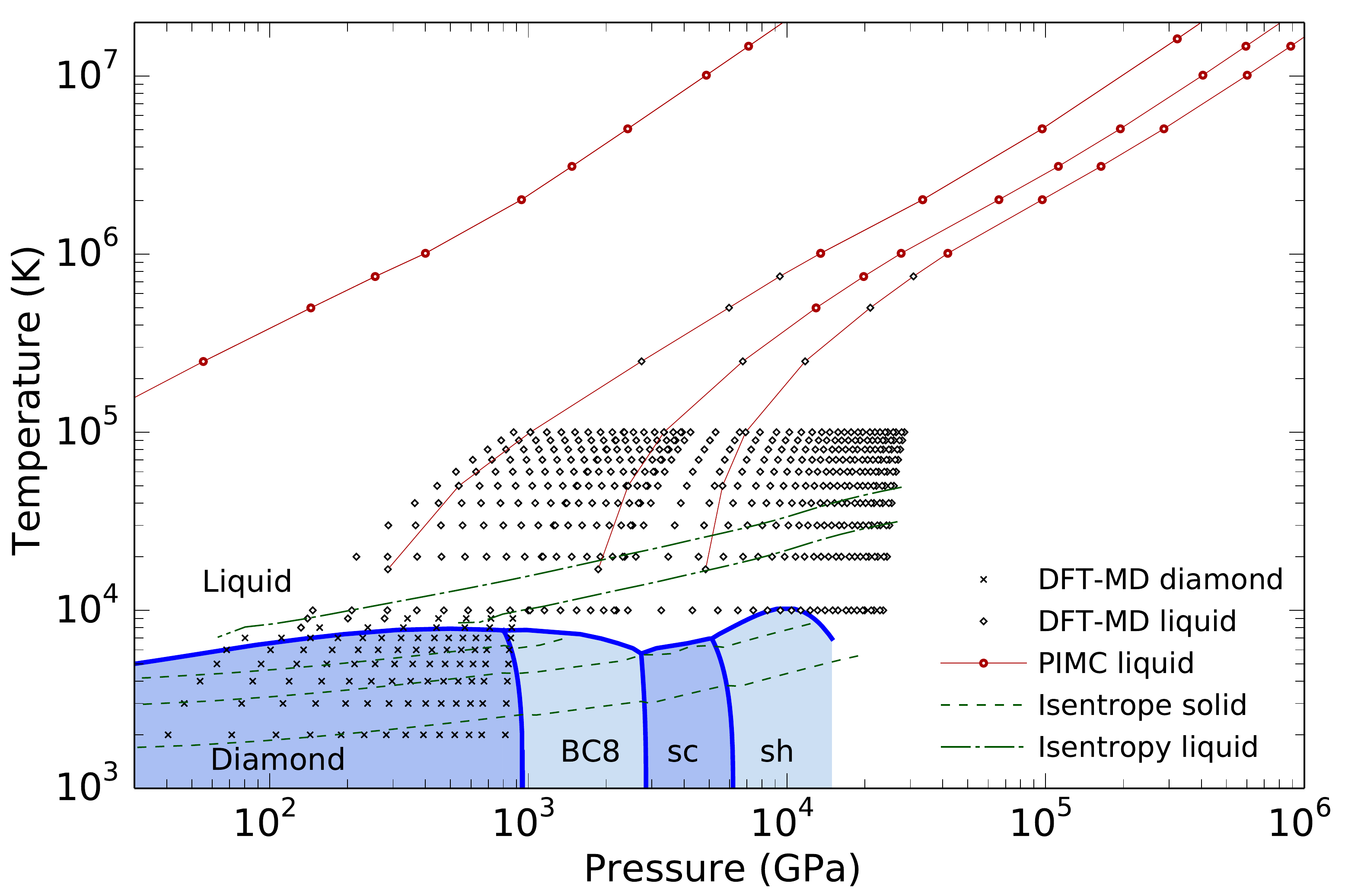}
\caption{Predicted C phase diagram (see subsequent sections) along with the DFT-MD (black open symbols for liquid, black crosses for solid) and PIMC (red open symbols) data we use to validate our EOS model. Note that the four isochores with densities 0.1 g/cc, 3.18 g/cc, 8.5 g/cc, and 11.18 g/cc (red lines) contain DFT-MD points for lower-$T$ and PIMC points for higher-$T$. Also shown are portions of five isentropes (green lines) computed from our EOS model with entropy values increasing from the bottom: 3.78 $k_{\rm B}$/atom, 5.49 $k_{\rm B}$/atom, 6.61 $k_{\rm B}$/atom, 12.56 $k_{\rm B}$/atom, and 12.72 $k_{\rm B}$/atom. Note that the 0.1 g/cc isochore (upper left-most red curve) exists throughout a range in which the EOS model of this work is {\it not} valid.} 
\label{pd_data}
\end{figure}

\subsection{DFT calculations}

With DFT methods, we perform calculations of cold curves, phonon densities of states, electronic excitation contributions to the free energy, and the EOS itself ($E$ and $P$ as functions of $\rho$ and $T$). We also present a limited number of calculations of ionic diffusivity, from which we can infer melt behavior. For all of these, we use the VASP code \cite{vasp}, together with projector augmented wave (PAW) pseudopotentials \cite{paw}. We use a "hard" PAW with a core radius of 1.1 bohr and 4 valence electrons and the generalized gradient approximation (GGA) of DFT with the Perdew-Burke-Ernzerhof (PBE) exchange correlation functional \cite{pbe}. The plane-wave cutoff is set to 1300 eV. 

For the phonon calculations, we use the primitive cell with a MP k-point grid of 40x40x40 for the simple cubic (sc), the simple hexagonal (sh) phases and for fcc. The sh is done with a $c/a$ ratio of 0.986 evaluated by performing cell optimization for several compressions. No appreciable variation in the $c/a$ ratio with volume was observed.
For the diamond and bc8 phases we used a 8 atom unit cell with a 20x20x20 k-points grid. For the sc phase, we compared with a 2 atom unit cell with a pmma spacegroup (tetragonal distorsion) \cite{MC} which was calculated with a 20x40x40 k-point grid.

For the molecular dynamics simulations, we use 64 atoms and perform Born-Oppenheimer MD (BOMD) within the NVT-ensemble with a Nos\'e-Hoover thermostat \cite{nose}. We use a time step of 0.75 fs in order to converge the internal energy and pressure to the desired accuracy. The electronic density is constructed from single-particle wave functions by sampling at the ($\frac{1}{4}$,$\frac{1}{4}$,$\frac{1}{4}$) point of the Brillouin zone. The electron occupation numbers are taken to be a Fermi-Dirac distribution set at the average temperature of the ions. For the different densities and temperatures, we use a sufficient number of bands such that we have at least 40 bands with occupation numbers smaller than 0.00001. The MD is run for 10000 steps with the last 5000 steps used for averaging the internal energy and the pressure.

\begin{figure}
\includegraphics[scale=0.40]{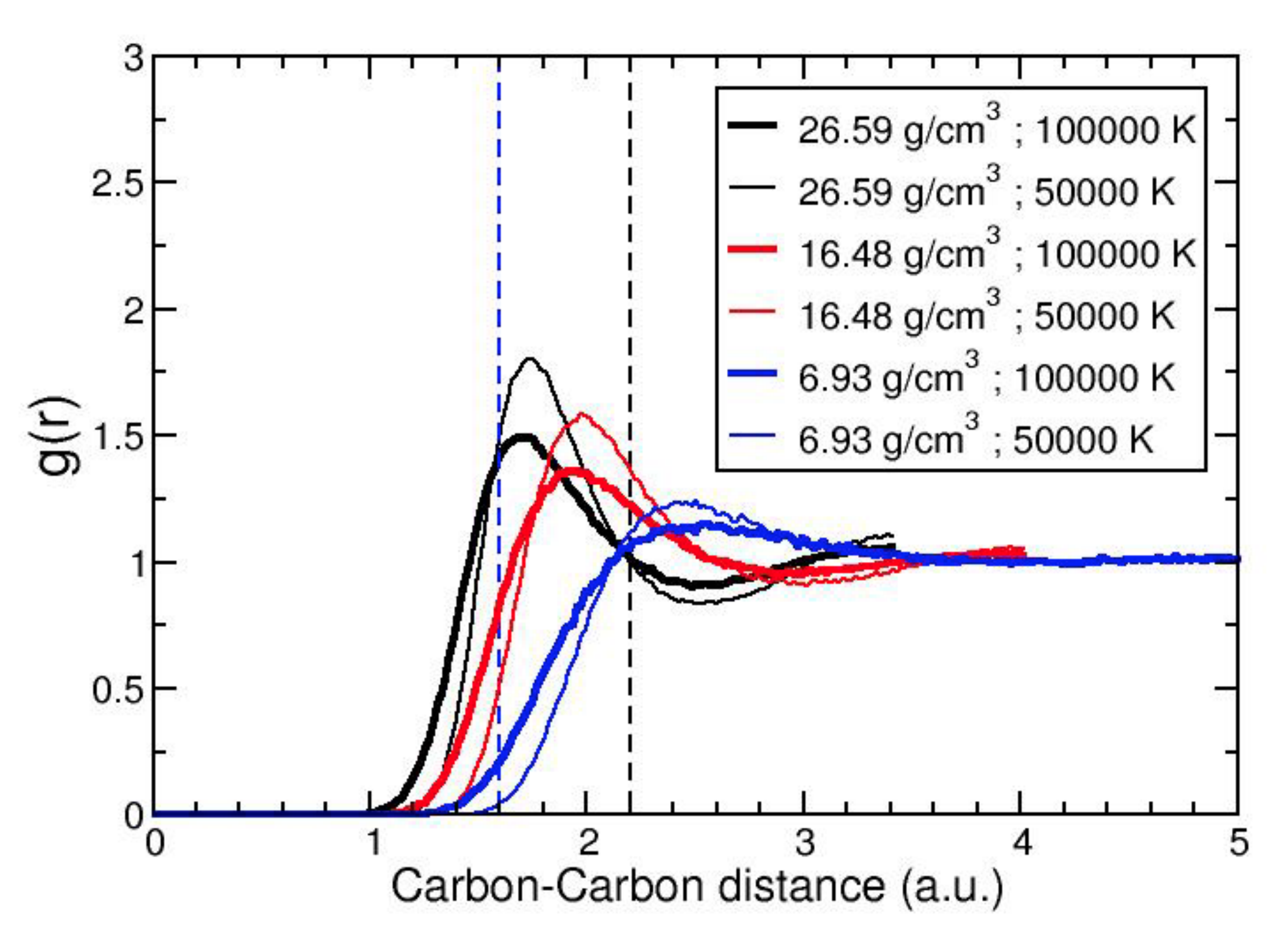}
\caption{Pair correlation function for high-pressure liquid carbon for $T= $T 50,000 K  and 100,000 K from PBE-DFT-MD. The black and blue dashed vertical lines are the diameter of the 1.1 and 0.8 bohr PAW potentials respectively.}
\label{hpgofr}
\end{figure}

The use of pseudopotentials under extreme conditions of pressure and temperature can be problematic. For high compressions or high temperatures the inter-atomic distances can become smaller than the diameter of the PAW sphere. In figure \ref{hpgofr}, we plot the pair distribution functions for the densities 26.59 g/cc, 16.48 g/cc and 6.93 g/cc and temperatures of 50000 K and 100000 K, where clearly the PAW spheres overlap significantly (the black and blue dashed vertical lines are the diameter of the 1.1 and 0.8 bohr PAW potentials respectively; see below). At higher temperatures still, the thermal excitations of the 1s electrons can no longer be neglected. This can be addressed by having all 6 electrons in the valence (for the current DFT-MD calculations, however, we find this not to be necessary for temperatures of 100,000 K or lower). At the highest compressions, the perturbation of the 1s orbital may be significant and can again be addressed by having all 6 electrons in the valence. 

\begin{figure}
\includegraphics[scale=0.40]{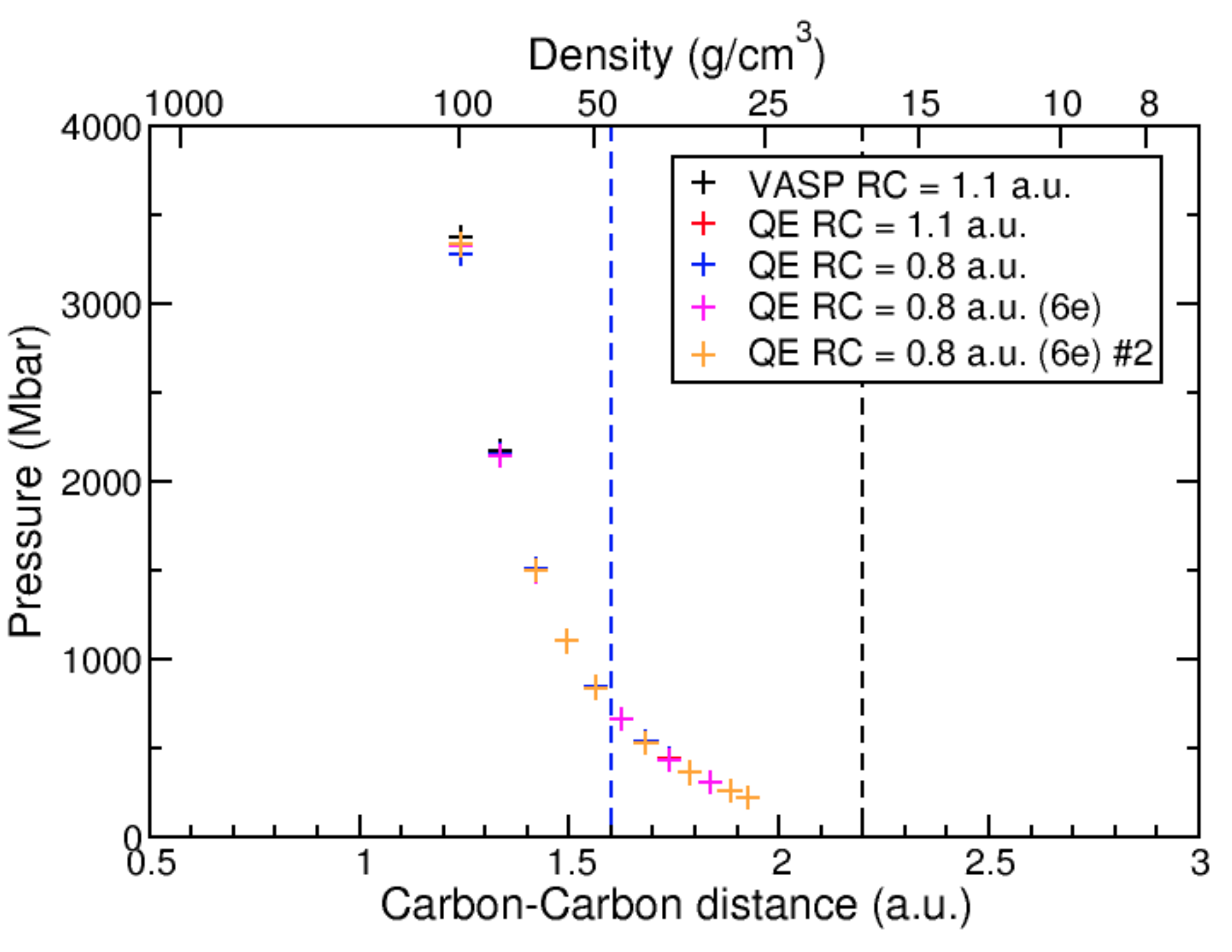}
\caption{Pressure (at $T= 0$ and for fixed ionic positions) versus inter-ionic distance for fcc carbon with various PAW potentials using PBE-DFT. The black and blue dashed vertical lines are the diameter of the 1.1 and 0.8 bohr PAW potentials respectively.}
\label{PAWtest}
\end{figure}

In order to assess the sensitivity of our results to the PAW potential chosen, we construct PAW potentials with different cutoff radii, which range from 1.1 to 0.8 bohr radii. The Vanderbilt projectors generation scheme is utilized \cite{Vanderbilt}. As the cutoff radii decrease, in order to maintain and optimize the PAW performance, we add additional partial waves for both $s$ and $p$ angular momenta. Different sets of reference energies are chosen, which are determined not to affect our computational results. Moreover, when the carbon-carbon distance is sufficiently small, the 1$s$ core state is also included in the PAW. For all cases tested, we find that a 5442~eV plane-wave cutoff energy is enough to reach convergence. For the PAW test, we use fcc cells with lattice constants chosen so that the carbon-carbon nearest neighbor distances are representative of the distances observed during the highest-$P$ cold curve and MD runs. These tests are performed with the Quantum-Espresso (QE) package \cite{pwscf}. Figure \ref{PAWtest} shows a comparison of the pressure obtained for this fcc cold curve with the different PAWs (for both VASP and QE).
We only observe appreciable deviations between the different PAWs for densities above 60 g/cc, corresponding to carbon-carbon distances of $\sim$ 1.3 bohr (0.7 $\AA$). 
The difference in the calculated pressure between the VASP $R_{\rm cut}$=1.1 bohr PAW potential and the 6 electron PAW is only 1\% up to densities of 100 g/cc. This corresponds to carbon-carbon distances of 1.1 bohr which is essentially the smallest carbon-carbon distances sampled during the highest temperature and highest density MD runs. However, this is no guarantee that the dynamics are not affected by the overlapping PAW spheres. In this sense, we submit that our DFT-MD EOS data for the very highest densities and temperatures reported here may be of lower accuracy than those for the more moderate conditions. Nevertheless, we stress that the bulk of our conclusions below are unaffected by this fact.

\subsection{PIMC calculations}
PIMC is the most accurate and efficient first-principles simulation
technique to study the equilibrium properties of quantum systems in
high temperature plasma states. Fermionic PIMC simulations have been
applied to study
hydrogen~\cite{PC94,Ma96,Mi99,MC00,MC01,Mi01,MG06,Hu2010,Hu2011},
helium~\cite{Mi06,Mi09}, hydrogen-helium mixtures~\cite{Mi05}, one
component plasmas~\cite{MP04,MP05} and most recently to simulate carbon and water
plasmas~\cite{Driver2012}. In PIMC simulations, electrons and nuclei
are treated equally as Feynman paths in a stochastic framework for
solving the full, finite-temperature, quantum many-body problem.  The
natural operator to work with in this context is the thermal density
matrix represented as a path integral in real space,
\begin{displaymath} 
\rho(\mathbf{R},\mathbf{R}';\beta) \equiv \left< {\bf R} | e^{-\beta\hat{\mathcal{H}}} | {\bf R}' \right>
\end{displaymath}
\begin{displaymath}
= \int \!\! \ldots \!\! \int {{d}}{\bf R}_{1}\:{{d}} {\bf R}_{2}\:\ldots\: {{d}} {\bf R}_{M-1}\; \rho({\bf R},{\bf R}_{1}; \Delta\tau ) \: \rho({\bf R}_{1},{\bf R}_{2};\Delta\tau) \ldots
\rho({\bf R}_{M-1},{\bf R}' ;\Delta\tau )
\end{displaymath}
\begin{equation}
= \int \limits_{{\bf R} \rightarrow {\bf R'} } \! \! \!  \mathcal{D}{\bf
R}(\tau) \exp\big\{-S[{\bf R}(\tau)]\big\}
\label{PIMCeq1}
\end{equation}
where $\mathcal{H}$ is a many-body Hamiltonian, $\beta=1/k_BT$ assumes the role of imaginary time, ${\bf R} = ({\bf r}_1,{\bf r}_2, \cdots {\bf r}_N)$, ${\bf r}_i$
is the position of the $i$th particle, and $\Delta\tau=\beta/M$  is the
step size in imaginary time intervals. $S[{\bf R}(\tau)]$ is the action, which determines the weight of every
path. The kinetic energy operator controls the diffusion of the paths
in imaginary time and keeps bead ${\bf R}_i$ at adjacent time slices
close together. A thermodynamic function corresponding to an operator,
$\hat{\mathcal{O}}$, can be derived from $\langle {\hat{\mathcal O}} \rangle
= {Tr}(\hat\rho\hat{\mathcal O}) / Z_c$, where
$Z_c={Tr}(\hat\rho)$ is the canonical partition function.

PIMC explicitly addresses all the physics of high-$T$ plasmas including
effects of bonding, ionization, exchange-correlation, and quantum
degeneracy on an equal footing, and thereby circumvents both the need to occupy single-particle states, and the need to employ exchange-correlation approximations inherent
in DFT. The Coulomb
interaction between electrons and nuclei is introduced using pair
density matrices derived using the eigenstates of the two-body Coulomb
problem~\cite{Pollock1988}. The periodic images are treated using an
optimized Ewald break-up~\cite{Natoli1995,MG06} applied to the pair
action~\cite{Pollock1984}.

Challenges with the PIMC formalism arise in fermionic simulations
because only antisymmetric eigenstates contribute to the partition
function. Those can be projected out by introducing an additional sum
over all permutations, $\rho_F({\bf R},{\bf R'},\beta)=1/N! \,
\sum_{\mathcal{P}} (-1)^{\mathcal{P}}\rho({\bf R},\mathcal{P} {\bf
  R'},\beta)$. Straightforward integration methods all lead to
unstable algorithms due to the \emph{fermion sign problem}, which is
the result of the near complete cancellation of positive and negative
contributions to the many-body density matrix.

A fixed-node approximation~\cite{Anderson1976,Reynolds1982} was
initially introduced to solve the sign problem for ground-state
quantum Monte Carlo calculations. For the finite-temperature PIMC
simulation, a restricted-path method was
developed~\cite{Ceperley1992}, where one only integrates over all
paths that do not cross the nodes of a {\it trial} density matrix,
$\rho_T({\bf R},{\bf R'};\beta)$, which must be available in analytic
form. The fermionic version of Eq.~\ref{PIMCeq1} then reads
\begingroup\makeatletter\def\f@size{9}\check@mathfonts
\begin{equation} 
\rho_F({\bf R},{\bf R}';\beta) = \frac{1}{N!}\sum_\mathcal{P}
(-1)^{\mathcal{P}} \!\!\!\!\!\!\!\!\!\!\!\!\!  
\int \limits_{ 
\begin{array}{c}
^{{\bf R} \rightarrow \mathcal{P} {\bf R'}}\\[-6pt]^{\rho_T({\bf R}(\tau),{\bf R'};\tau))>0}
\end{array} }
\!\!\!\!\!\!\!\!\!\!\!\!\!  \mathcal{D}{\bf R}(\tau) \exp\big\{-S[{\bf
R}(\tau)]\big\}. 
\end{equation}
\endgroup

In addition to weighting all paths according to their action and
summing over permutations, one must also check whether the sign
of the trial density matrix has remained positive during every Monte
Carlo move. If a sign change occurs anywhere along the new path, the
proposed move is rejected. The ${\bf R'}$ argument in
the nodal restriction plays a special role because the sign of the
trial density matrix at all other time slices depends on it.

The restricted-path method provides the exact answer if the nodes from
the exact many-body density matrix are used. Since exact nodes are not
known for interacting systems, one must work with approximate
nodes. However, within any given set of nodes, the restricted path
method will obtain the best possible solution that includes all
interaction and correlation effects. The PIMC method used here employs
a free-particle nodal structure, which has been shown to be sufficient
for carbon at temperatures where atoms still have occupied 1s states,
but partially occupied 2s states ($T \ge 2.5\times10^5$
K)~\cite{Driver2012}.

Using the restricted-path, free-particle nodal framework, we use
all-electron PIMC to compute equations of state for carbon at
densities of 0.1, 3.18, 8.5, and 11.18 g$\,$cm$^{-3}$ over a temperature
range of 10$^5$--10$^9$ K (though we present data pertaining to all four isochores in the Supplementary Material, we stress that 0.1 g/cc is below the low-density limit of applicability of the EOS model described below) \cite{lowrho}. A sufficiently small time step associated
with PIMC path slices is determined by converging total energy as a
function of time step until it changes by less than 0.2\%. We use
a time step of 0.0078125 Ha$^{-1}$ for temperatures below
5$\times10^6$ K and, for higher temperatures, the time step decreases
as $1/T$ while keeping at least four time slices in the path
integral. In order to minimize finite size errors, the total energy was
converged to better than 0.2\% for a 24 atom cubic cell.

\section{Construction and discussion of EOS model}
We make the fundamental assumption that the Helmholtz free enegy, $F= E - TS$, of each phase can be decomposed into the following terms:
\begin{equation}
F(V,T)= E_{\rm cold}(V) + F_{\rm ion}(V,T) + F_{\rm elec}(V,T),
\label{breakdown}
\end{equation}
where $F_{\rm ion}$ and $F_{\rm elec}$ represent the contributions from ionic and electronic excitations, respectively. While the identification of a temperature-independent piece ($E_{\rm cold}$) is always possible by virtue of the fact that it is merely definitional, the separation of the thermal part into decoupled ionic and electronic pieces is a major assumption. In a practical sense, this assumption is at least a reasonable starting point since for solids at lower-$T$, it has been shown to be accurate \cite{Wallace,Correa}, and for the liquid at very high temperature (ideal gas limit) it is also trivially satisfied because interactions between particles are of no importance. It is not necessarily justified for intermediate temperatures, but we shall invoke the approximation at any rate, as has been done in the past \cite{Correa,Benedict,Correa13}and take comfort in the fact that the ab initio EOS data to which we fit in the liquid (obtained from DFT-MD and PIMC) does {\it not} make this assumption.

\subsection{solid phases}
For the solid phases we consider diamond, BC8, simple cubic (sc), and simple hexagonal (sh), we take $E_{\rm cold}(V)$ to be the internal energy of the perfect crystalline lattice (with motionless ions) at $T= 0$ \cite{caveatcold} with the electrons in their ground state. We fit this term to the DFT data for the internal energy of the perfect lattice using the Vinet equation of state \cite{Vinet},
\begin{displaymath}
E_{\rm cold}(V)= \phi_{0} + \frac{4V_{0}B_{0}}{(B' - 1)^{2}}[1 - (1 + X)\exp(-X)],
\end{displaymath}
where
\begin{equation}\label{Vinet} 
X= \frac{3}{2}(B' -1)[(V/V_{0})^{1/3} - 1].
\end{equation}
\begin{figure}
\includegraphics[scale=0.40]{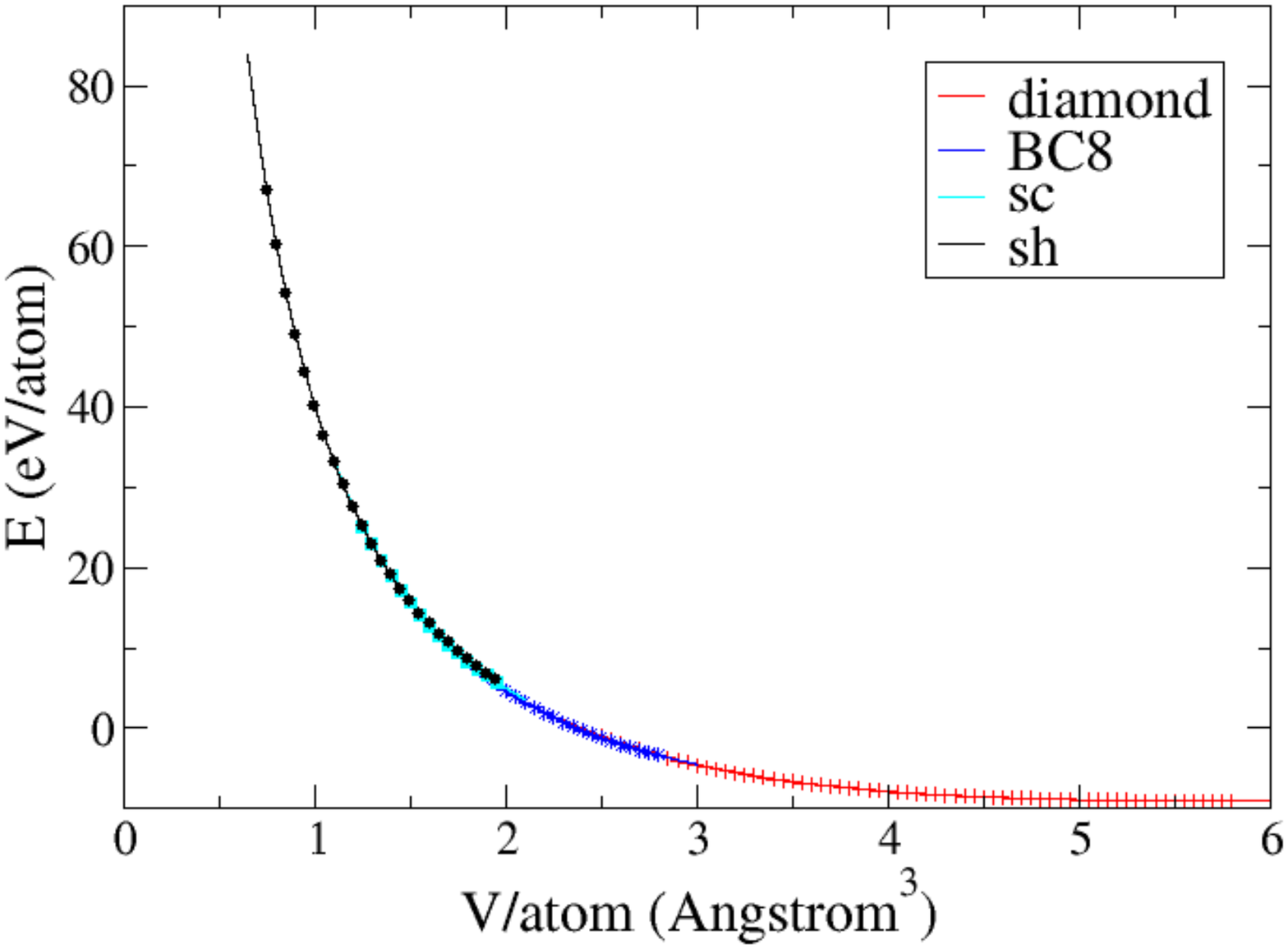}
\includegraphics[scale=0.40]{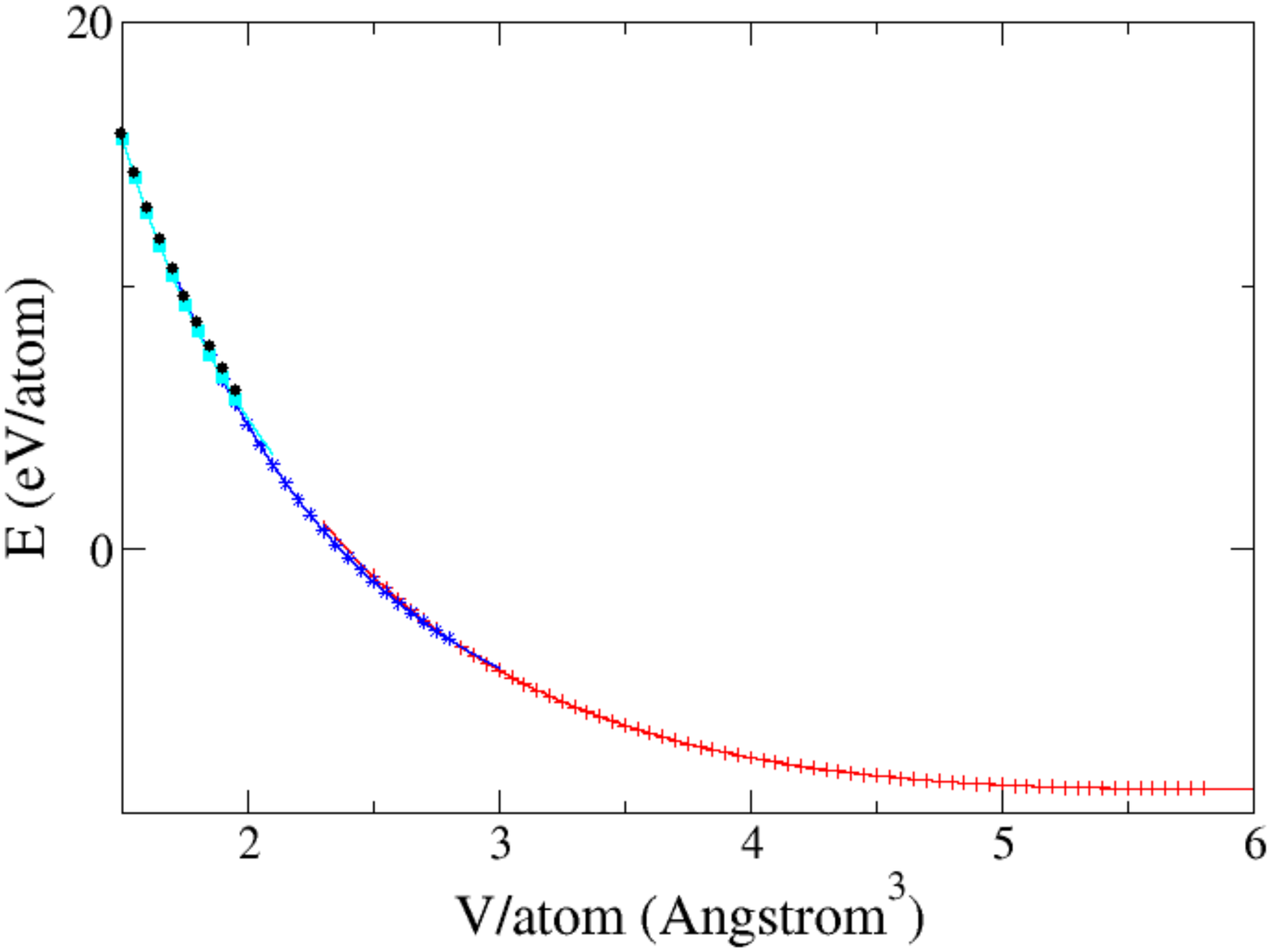}
\caption{Internal energy at $T$= 0 (and for fixed ions) in eV/atom versus $V$/atom in $\AA^{3}$ for various phases of C. Points indicate the results of PBE-DFT calculations, while the thin curves are fits to these data used in our EOS model. The phases represented are: diamond (red), BC8 (blue), sc (cyan), and sh (black). (a) shows the whole range over which the EOS model is constructed, while (b) shows a closer view of the range between $V= $ 2 and 5 $\AA^{3}$/atom. 
}
\label{colds}
\end{figure}
Here, $B_{0}$ is the bulk modulus at the volume $V_{0}$, $B'$ is the pressure-derivative of the bulk modulus at $V_{0}$, and $\phi_{0}$ is the internal energy at $V_{0}$. Fig. \ref{colds} shows the computed cold curves of the various phases, together with our fits to them, using Eq.\ref{Vinet}. The cold curve parameters for each phase ($\phi_{0}$, $B_{0}$, $B'$, and $V_{0}$) are given in Table 1, along with the other phase-dependent parameters we discuss below. Our (PBE) GGA-DFT results for the solid cold curves are very similar to those of Ref.\cite{MC}, but there are differences which result presumably from the use of different pseudopotentials and associated choices for the plane wave energy cutoff in the determination of the internal energy. Though small,  these differences give rise to rather pronounced changes in the predicted phase transition pressures (BC8 $\to$ sc $\to$ sh) when compared to the phase diagram presented in Ref.\cite{MC}. This will be discussed more below, after the thermal components of the free energies are considered and our phase diagram is presented. At this point, we note that we have checked that our inferred {\it cold} transition pressures, determined from the two-phase Maxwell construction \cite{Wallace}, should be inaccurate by no more than a few percent due to our use of the Vinet fitting form, Eq.\ref{Vinet} (though we in no way imply that PBE GGA-DFT is necessarily this accurate for predicting these transition pressures at such high compressions).

\begin{figure}
\includegraphics[scale=0.40]{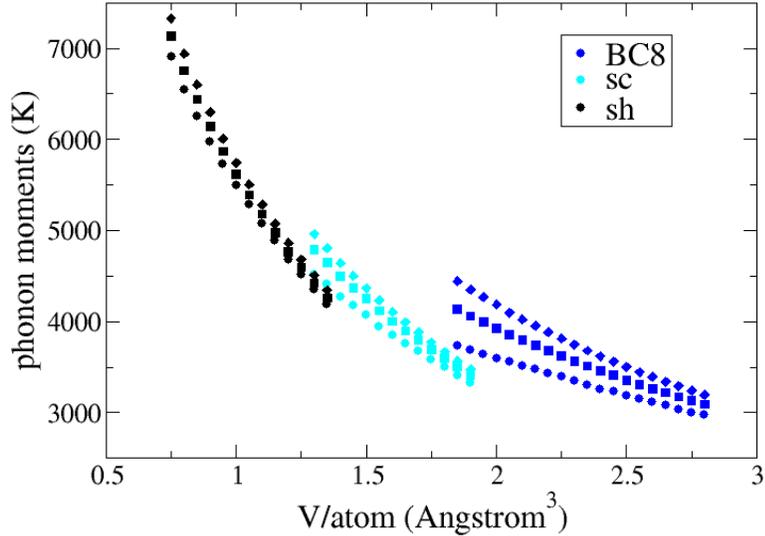}
\caption{Moments of the phonon densities of states, $\theta_{0}$ (circles), $\theta_{1}$ (squares), $\theta_{2}$ (diamonds) in Kelvins versus $V$/atom in $\AA^{3}$ for BC8 (blue), sc (cyan), and sh (black) phases as computed with linear response methods using PBE-DFT.}
\label{thetas}
\end{figure}

For $F_{\rm ion}(V,T)$, we use the ion excitation free energies of the various lattices computed for  cases in which the electrons are in their instantaneous ground state (for each ionic configuration). In other words, we invoke the Born-Oppenheimer approximation. In addition, the quasiharmonic approximation is assumed for each phase. We investigated anharmonicity for diamond and BC8 phases in Ref.\cite{Correa} and showed its effects to be small; the effects of anharmonicity in sc and sh phases await further study. Given this quasiharmonic description, we use double-Debye models \cite{Correa,Correa13} for the diamond, BC8, and sc phases, and a regular single-Debye model for the sh phase. All are fit to the $V$-dependent phonon densities of states (PDOSs) for these various phases computed with DFT using a linear response approach. 
In the double-Debye prescription, the PDOS for a given volume is presumed to consist of two overlapping features, each of which is of the form: $a\varepsilon^{2}$ for $\varepsilon \leq \theta$ and 0 for $\varepsilon > \theta$, where the two unequal $\theta(V)$ functions are related to various moments of the PDOS \cite{Correa}. Since the double-Debye model allows for two different values of $\theta$ (say, $\theta_{\rm A}$ and $\theta_{\rm B}$) for each $V$, it is useful in cases where the high-$T$ phonon moment, $k_{\rm B}\theta_{0}(V)= e^{1/3}\exp\left[\int_{0}^{\infty}d\varepsilon \ln (\varepsilon) D_{V}(\varepsilon)\right]$, is rather different from the low-$T$ zero-point moment, $k_{\rm B}\theta_{1}(V) = 4/3\int_{0}^{\infty}d\varepsilon \varepsilon D_{V}(\varepsilon)$. This is certainly the case for the diamond and BC8 phases, especially at high compressions, as discussed in detail in Ref.\cite{Correa}. Our calculations for the sh phase shows that $\theta_{0}$ and $\theta_{1}$ are very nearly equal throughout the predicted range of its stability. Thus, we use a single-Debye free energy for the sh ion-thermal term to reduce the total number of fitting parameters. Fig. \ref{thetas} shows the three phonon moments, $\theta_{0}, \theta_{1}$, and $\theta_{2}$ ($k_{\rm B}\theta_{2}(V)= \sqrt{\frac{5}{3}\int\varepsilon^{2}D_{V}(\varepsilon)d\varepsilon})$, as functions of $V$ for BC8, sc, and sh phases as determined from our computed PDOSs. Note that the $\theta_{0}(V)$ for the different phases nearly coincide in the neighborhood of the volumes where their cold curves cross. For this reason, the $T$ vs. $P$ phase lines are rather vertical in nature (see below). It is from these moments that we derive (for the diamond, BC8, and sc phases where the double-Debye model is used) the two Debye temperatures, $\theta_{\rm A}$ and $\theta_{\rm B}$, by solving the system of equations given as Eqs.9 - 12 of Ref. \cite{Correa}. 
For all $\theta_{\rm A,B}(V)$ functions for the various phases, we assume the $V$-dependence:
\begin{equation}
\theta(V)= \theta^{0}\left(\frac{V}{V_{p}}\right)^{-B}\exp\left[A(V_{p} - V)\right],
\label{gamma}
\end{equation}
where $\theta^{0}$ is the value at a reference volume $V_{p}$. 
This arises from the assumption that the ion-thermal Gr\"uneisen parameter for $\theta(V)$ is equal to $\gamma= AV + B$ \cite{Correa,Benedict}. The relevant ion-thermal parameters ($\theta^{0}$, $A$, $B$, and $V_{p}$) which determine the various Debye temperatures for each phase are presented in Table 1. In terms of $\theta_{\rm A}$, $\theta_{\rm B}$, and $\theta_{1}$, the double-Debye free energy is written as
\begin{equation}
F_{\rm ion}(V,T)= \left[\frac{\theta_{\rm B}(V) - \theta_{1}(V)}{\theta_{\rm B}(V) - \theta_{A}(V)}\right]F_{\rm A}(V,T) + \left[\frac{\theta_{1}(V) - \theta_{\rm A}(V)}{\theta_{\rm B}(V) - \theta_{\rm A}(V)}\right]F_{\rm B}(V,T),
\end{equation}
where $F_{\rm A}$ and $F_{\rm B}$ are the single-Debye free energies for the individual A and B peaks associated with a given PDOS.

The electronic excitation term for each solid phase, $F_{\rm elec}(V,T)$, is taken be of the form, 
\begin{equation}
F_{\rm elec}(V,T)= -\frac{1}{2}\alpha (V)T^{2}k_{\rm B}/{\rm atom},
\label{Fesol}
\end{equation}
which is motivated by a performing a Sommerfeld expansion of the electronic excitation free energy, assuming $T/T_{\rm Fermi}$ to be a small parameter \cite{Wallace}. This contribution is extracted from PBE GGA-DFT calculations in which electronic occupancies are constrained by a Fermi-Dirac distribution with temperature $T$, but for ions {\it fixed} in their lattice positions. As discussed in Ref.\cite{Correa}, this term is essentially zero for the diamond phase throughout its stability field, and is exceedingly small (though non-zero) for the BC8 phase as well \cite{caveatelec}. For this reason, electronic excitation contributions were intentionally neglected by those authors in their solid-phase EOS models. Here we choose to include these terms; their values for the sc and sh phases are somewhat larger than those of BC8, because the electronic density of states near $E_{\rm Fermi}$ generally increases with compression. Still, their inclusion has a rather small effect on EOS and the resulting phase lines. We fit these small DFT-based terms by assuming $\alpha(V)= \alpha_{0}(V/V_{e})^{\kappa}$ \cite{Wallace,Correa,Benedict}; our choices for $\alpha_{0}$, $\kappa$, and $V_{e}$ are presented in Table 1 for each phase. Note that there are nonzero values for $\alpha_0$ which appear for the diamond and BC8 phases in this table; these result from earlier extractions of anharmonic ion-thermal free energy contributions \cite{Correa} (which happen also to have a $T^{2}$ dependence), and are {\it not} due to electronic excitations \cite{caveat_anh}. 

\begin{figure}
\includegraphics[scale=0.40]{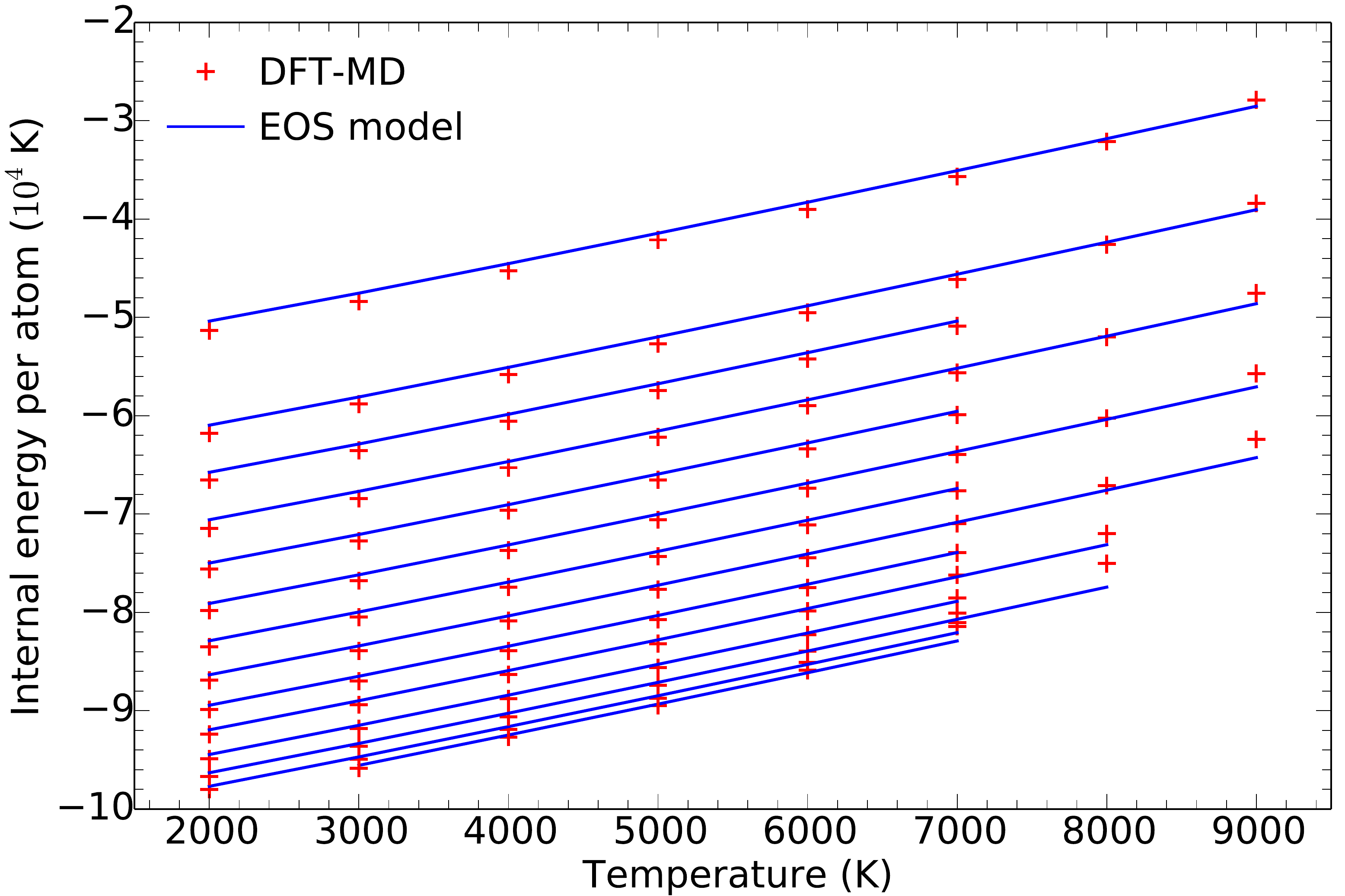}
\includegraphics[scale=0.40]{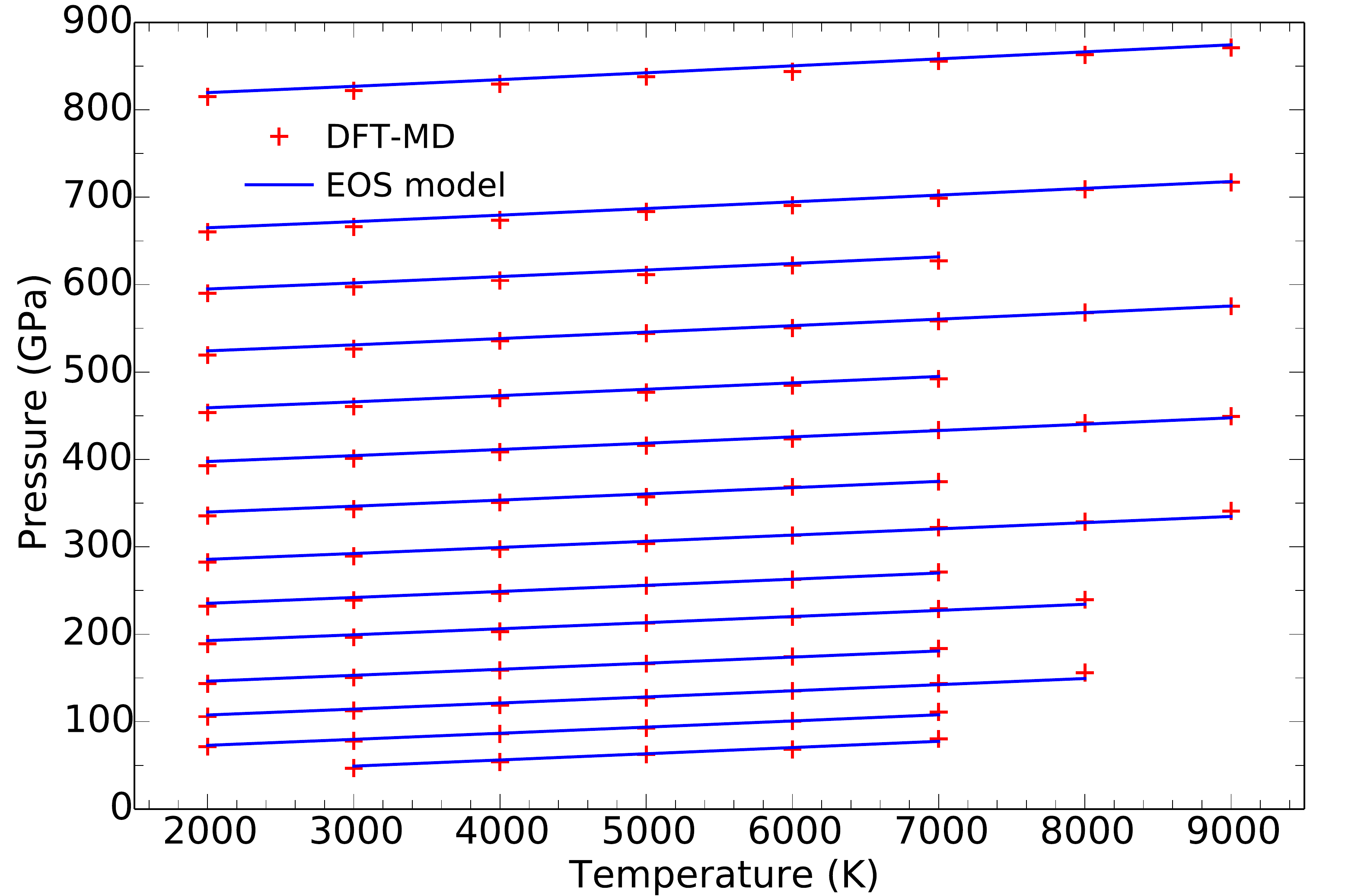}
\caption{(a) Internal energy (K/atom) isochores versus $T$ (K) for the diamond phase. (b) Pressure (GPa) isochores versus $T$ (K) for the diamond phase. Red points are the results of DFT-MD, blue lines are the results of our EOS model. In both plots, $V$ ranges from 3.0683 $\AA^{3}$/atom to 5.3903 $\AA^{3}$/atom.
}
\label{diamond}
\end{figure}
 
In order to check the validity of our solid-phase EOS model for C, we compare to our DFT-MD results for pressure and internal energy in the diamond phase. These results exist throughout a range where diamond is predicted to be the stable phase ($3.0 \AA^{3}/{\rm atom} < V < 5.6 \AA^{3}/{\rm atom}$, and $2000 {\rm K} < T < 9000 {\rm K}$). Fig. \ref{diamond} shows that the model agrees with the diamond-phase DFT-MD quite well, even though these data, per se, were not used in the fitting (rather, only cold curves, PDOSs, and electronic DOSs were used). Various factors contribute to the slight disagreement between the model and the DFT-MD data: 1. The model is built from the diamond cold curve as calculated for an effectively infinite system (affected by a large number of ${\bf k}$-points), while the DFT-MD was performed with a limited number of atoms (64) in a supercell and a single ${\bf k}$-point. 2. The ion-thermal term of the diamond free energy model is taken to be that of Ref.\cite{Correa}; the phonon calculations therein were performed with different pseudopotentials and different plane wave cutoff energies. This is likely to be a very small contributor to the EOS differences, however, for the relevant moments of the PDOSs at various volumes are strikingly similar to those resulting from the computational framework adopted in this work.

\begin{table}
\begin{center}
\begin{tabular}{| c | c | c | c | c |}
\hline
parameter & diamond & BC8 & sc & sh \\ \hline
$V_{0}$ & 5.7034 & 6.242 & 7.9899 & 9.6061  \\ \hline
$B_{0}$ & 432.4 & 221.2 & 59.09 & 22.12  \\ \hline
$B'$ & 3.793 & 4.697 & 5.763 & 6.495  \\ \hline
$\phi_{0}$ & -9.066 & -8.705 & -7.525 & -6.5  \\ \hline\hline\hline
$V_{p}$ & 5.571 & 3.176 & 2.658 & 1.35  \\ \hline
$\theta^{0}_{A}$ & 1887.8 & 1961.9 & 2089.8 & 4183.8  \\ \hline
$A_{A}$ & -0.316 & 0.0 & 0.0 & 0.4354 \\ \hline
$B_{A}$ & 0.913 & 0.0 & 0.212 & 0.4034  \\ \hline
$\theta^{0}_{B}$ & 1887.8 & 3176.3 & 2961.3 & 4183.8  \\ \hline
$A_{B}$ & 0.168 & 0.156 & 0.0 & 0.4354  \\ \hline
$B_{B}$ & 0.429 & 0.532 & 0.817 & 0.4034  \\ \hline
$\theta^{0}_{1}$ & 1887.8 & 2800.6 & 2328.3 & 4183.8  \\ \hline
$A_{1}$ & 0.0846 & 0.112 & 0.369 & 0.4354  \\ \hline
$B_{1}$ & 0.499 & 0.449 & 0.302 & 0.4034  \\ \hline\hline\hline
$V_{e}$ & 5.785 & 5.077 & 1.0 & 1.0  \\ \hline
$\alpha_{0}$ & 3.79e-5 & 5.5e-5 & 1.37e-5 & 1.58e-5  \\ \hline
$\kappa$ & 0.0 & 0.0 & 0.637 & 0.81  \\ \hline
\hline
\end{tabular}
\caption{Phase-dependent EOS model parameters for the solid phases of our multiphase C EOS. The upper segment of the Table concerns cold curve parameters, the middle segment quasiharmonic ion-thermal parameters, and the lower segment electron-thermal/anharmonic-ion-thermal parameters. All volumes ($V$) are in $\AA^{3}$/atom, $B_{0}$ is in GPa, $B'$ is unitless, $\phi_{0}$ is in eV/atom, all characteristic temperatures ($\theta$) are in Kelvins, $A$-parameters are in $\AA^{-3}$, $B$-parameters are unitless, $\alpha_{0}$ is in Kelvins$^{-1}$, and $\kappa$ is unitless.}
\end{center}
\end{table}

\subsection{liquid phase}
For liquid C, we assume a model much like that of our solid phases, consisting of separate cold, ion-thermal, and electron-thermal terms as shown in Eq.\ref{breakdown}. Here, however, there is no natural way to determine them all independently, because the liquid cannot be described as a perturbation away from a fixed configuration of ions. Nevertheless, we assume that Eq.\ref{breakdown} holds for the liquid and base our model on the liquid free energy assumed in Ref.\cite{Correa}, though with some important differences. In this approach, the cold piece is once again assumed to be of the Vinet form \cite{Vinet} (Eq.\ref{Vinet}), though augmented with a few so-called break-points which introduce bends in $E_{\rm cold}(V)$ over localized regions in $V$ \cite{Correa,bp}; additional bends at small $V$ are included to improve agreement with our liquid DFT-MD data performed at densities far higher than those considered in the work of Ref.\cite{Correa}. The ion-thermal term at lower-$T$ (up to $\sim$ twice $T_{\rm melt}$) is treated in the scheme of Chisolm and Wallace \cite{CW}, which is an effective Mie-Gr\"uneisen model for the liquid with a $V$-dependent characteristic temperature, $\theta (V)$, which we assume to have the form of Eq.\ref{gamma}. This model was used before for liquid C \cite{Correa}, and is constrained to have an ion-thermal specific heat, $C_{V}^{\rm ion}= 3k_{\rm B}$/atom; this is reasonable considering the ($T < $ 20,000 K) DFT-MD data that was provided in order to inform the liquid C EOS model in that earlier work \cite{Correa}. The contribution to the liquid C EOS from electronic excitations is predicted to be quite small for the lower temperatures, even though liquid C is predicted to be metallic \cite{Correa}. Nevertheless, we use the average-atom-in-jellium DFT model,  PURGATORIO \cite{PURGATORIO}, for the electron-thermal model of the liquid throughout the entire field of stability of the phase. As expected, its effects are pronounced at high-$T$; we discuss them below. The comparison of the resulting liquid C EOS model to our DFT-MD EOS data taken over a wide range of compression is shown below, after the high-$T$ liquid is addressed.

\subsubsection{The high-$T$ liquid: approach to the ideal gas}
Since we now have first-principles electronic structure data (DFT-MD and PIMC) for liquid C at {\it far} higher $T$ than we had previously \cite{Correa}, we are able to test various assumptions regarding the manner in which $C_{V}^{\rm ion}$ evolves from its Dulong-Petit value ($3k_{\rm B}$ per atom) to the ideal gas limit ($3k_{\rm B}/2$ per atom). To frame this discussion, we use two different high-$T$ additions to the liquid ion-thermal free energy, the predictions of which we will compare to these data: 1. A variant of the the Cowan model \cite{Cowan,MGextension}, and 2. The cell model \cite{Correa13}. Both models essentially satisfy the limit:
\begin{equation}
C_{V}^{\rm ion} \to \frac{3}{2}k_{\rm B} + \eta\left[\frac{T_{\rm ref}(V)}{T}\right]^{\nu}
\label{Cv}
\end{equation}
{\it for sufficiently high $T$}, where $T_{\rm ref}(V)$ is a $V$-dependent reference temperature.
For the Cowan model, the exponent $\nu$= 1/3; for the cell model, $\nu$= 2. Each have $C_{\rm V}^{\rm ion}= 3k_{\rm B}$  for $T$ well below $T_{\rm M}$ and $C_{\rm V}^{\rm ion}= 3k_{\rm B}/2$ as $T \to \infty$, but the rate at which the high-$T$ limit is approached is very different. In our version of the Cowan model, we choose the specific form of $C_{V}^{\rm ion}$ to be 
\begin{equation}
C_{V}^{\rm ion}= \frac{3}{2}k_{\rm B} + \frac{3}{2}k_{\rm B}\left[\frac{T_{\rm M}(V)}{T}\right]^{1/3},
\label{CvCowan}
\end{equation}
for all $T > T_{\rm M}(V)$.
Here, $T_{\rm M}(V)$ is constrained to satisfy the Lindemann law \cite{Lindemann} with the ion-thermal Gr\"uneisen parameter of the liquid EOS model of Ref.\cite{Correa} and we choose $T_{\rm M}(V_{0})$= 20,000 K at the ambient volume, $V_{0}$ \cite{caveatTm}. The ion-thermal free energy is then obtained by integrating this $C_{V}^{\rm ion}(T)$ appropriately from $T_{\rm M}$ to $T$ \cite{MGextension}, giving
\begin{displaymath}
F_{\rm ion}(V,T)= -\frac{3}{4}k_{\rm B}T_{\rm M}(V) + \frac{27}{4}k_{\rm B}T_{\rm M}(V)^{1/3}T^{2/3}
\end{displaymath}
\begin{equation}
-\frac{3}{2}k_{\rm B}T\log\left[\frac{T}{T_{\rm M}(V)}\right] - 3k_{\rm B}T\log\left[\frac{T_{\rm M}(V)}{\theta(V)}\right] - 6k_{\rm B}T,
\label{MG}
\end{equation}
where $\theta(V)= e^{-1/3}\theta_{\rm 0}(V)$ is the characteristic temperature of the liquid model in Ref.\cite{Correa}.
This expression is used for $T > T_{\rm M}$; for $T \leq T_{\rm M}$, the Mie-Gr\"uneisen form is used: $F_{\rm ion}(V,T)= 3k_{\rm B}\log[\theta(V)/T]$. In the cell model, the decay to the ideal gas form is affected by an additive (rather than piecewise) correction, derived by appealing to the classical approximation to the partition function for an ensemble of Einstein oscillators with each particle strictly confined \cite{Correa13} to a radius $R$. The resulting free energy, giving rise to $C_{V}^{\rm ion}$ approximately (for $T \gg T_{\rm M}$) equal to that of Eq.\ref{Cv} with $\nu= 2$, is 
\begin{equation}
F_{\rm ion}(V,T)= 3k_{\rm B}T\log\left[\frac{\theta}{T}\right] - k_{\rm B}T\log\left[{\rm erf}\left(\sqrt{\frac{T_{\rm M}}{T}}\right) - \frac{2}{\sqrt{\pi}}\sqrt{\frac{T_{\rm M}}{T}}{\rm e}^{-T_{\rm M}/T}\right],
\label{cell}
\end{equation}
where here, $k_{\rm B}T_{\rm M}(V)= m k_{\rm B}^{2}\theta^{2}R^{2}/2\hbar^{2}$ \cite{Lindemann}. In the simplest prescription, the confining radius is related to the volume per atom by $4\pi R^{3}/3= V$. This gives rise to a slightly incorrect value for the entropy at $T= \infty$, due to the explicit neglect of permutations between particles which are confined to individual cells. The correct high-$T$ limit of the entropy can be recovered by instead choosing the relationship, $4\pi R^{3}/3= Ve$ (where $\ln(e)= 1$) \cite{Correa13}. However, we find below that for fitting to $E$ and $P$ within the range of $(\rho,T)$ considered here, the choice of $4\pi R^{3}/3= V$ is preferred. In our practical implementation of the cell model we replace the first term in Eq.\ref{cell} (Mie-Gr\"uneisen) by a Debye model free energy with the identical characteristic temperature, $\theta$, to capture the thermodynamics of quantum ions at the lower temperatures (however subtle that distinction may be for liquid C).

\begin{figure}
\includegraphics[scale=0.40]{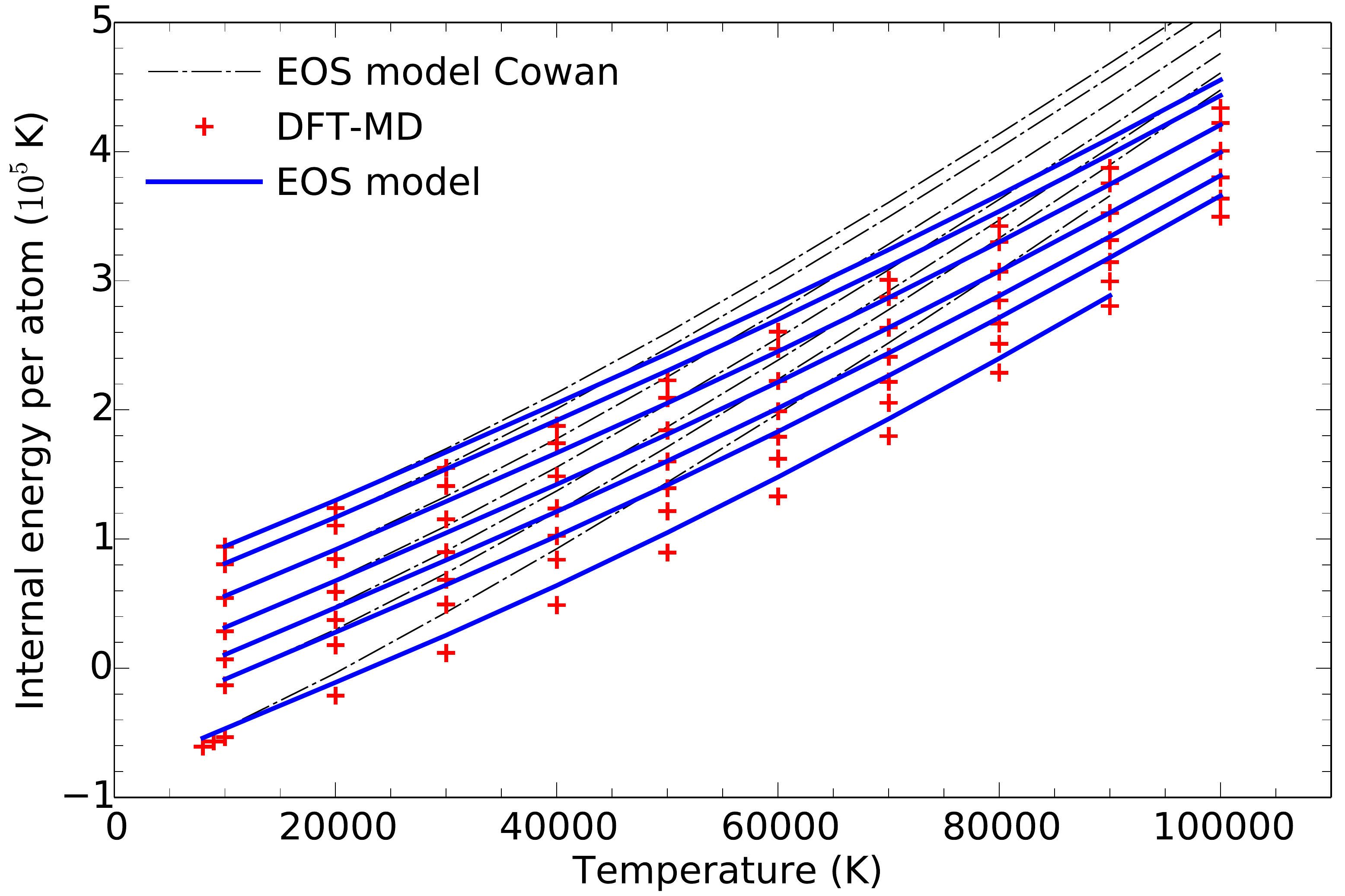}
\caption{Internal energy (K/atom) isochores versus $T$ (K) for the liquid as computed by DFT-MD (red points) and two variants of our liquid carbon EOS model; $F_{\rm ion}$= Cowan (black dashed-dotted lines), $F_{\rm ion}$= cell (blue lines). $V$ ranges from 2.05610 $\AA^{3}$/atom to 5.39030 $\AA^{3}$/atom. Note that this plot contains roughly one-half of the isochores in this density range for which we produced DFT-MD data.}
\label{ESH}
\end{figure}

Fig. \ref{ESH} shows internal energy, $E$, versus $T$ for liquid C on a densely-spaced grid of isochores for densities: 3 g/cc $< \rho <$ 12 g/cc. Red points are the results of DFT-MD extending up to $T=$ 100,000 K as described in Section II A. Black dashed-dotted lines are the predictions of a liquid C EOS model for these same $(\rho,T)$ states, which has been constructed exactly along the lines of Ref.\cite{Correa}, but with two important additions: 1. $F_{\rm ion}$ includes the Cowan model correction as described above (Eq.\ref{MG}) and in Ref.\cite{MGextension}; 2. The PURGATORIO model for C is used for $F_{\rm elec}$. Note that while the agreement is good for $T <$ 20,000 K, as ensured by the fitting scheme of Ref.\cite{Correa}, the slope of $E$ vs. $T$ for higher $T$ is much larger in the model, indicating that the model's $C_{V}$ is too large above 20,000 K. Also shown in Fig. \ref{ESH} is the results of an otherwise identical model (blue lines), but  with $F_{\rm ion}$ chosen to be the cell model; once again, PURGATORIO is used for $F_{\rm elec}$. Red points and blue lines in this plot are in far better accord. Assuming the electronic excitation contribution to be that of the PURGATORIO model, this comparison is showing that a faster decay of $C_{V}^{\rm ion}$ is greatly favored over the slower decay previously assumed \cite{MGextension}. This is at least vaguely troubling, since many EOS tables in the Livermore Equation of State (LEOS) database are constructed using the paradigm of QEOS \cite{QEOS}, which assumes the Cowan model for $F_{\rm ion}$. Of course, our present comparison is just for carbon, and is also for a somewhat restricted range of $\rho$ and $T$. Furthermore, it is entirely possible that 1. This conclusion depends on the electron-thermal model chosen, here PURGATORIO rather than the choice outlined in QEOS, average-atom Thomas-Fermi \cite{QEOS,Hora}, and 2. The DFT-MD data to which we compare is somehow lacking in fidelity at the higher-$T$. 
\begin{figure}
\includegraphics[scale=0.40]{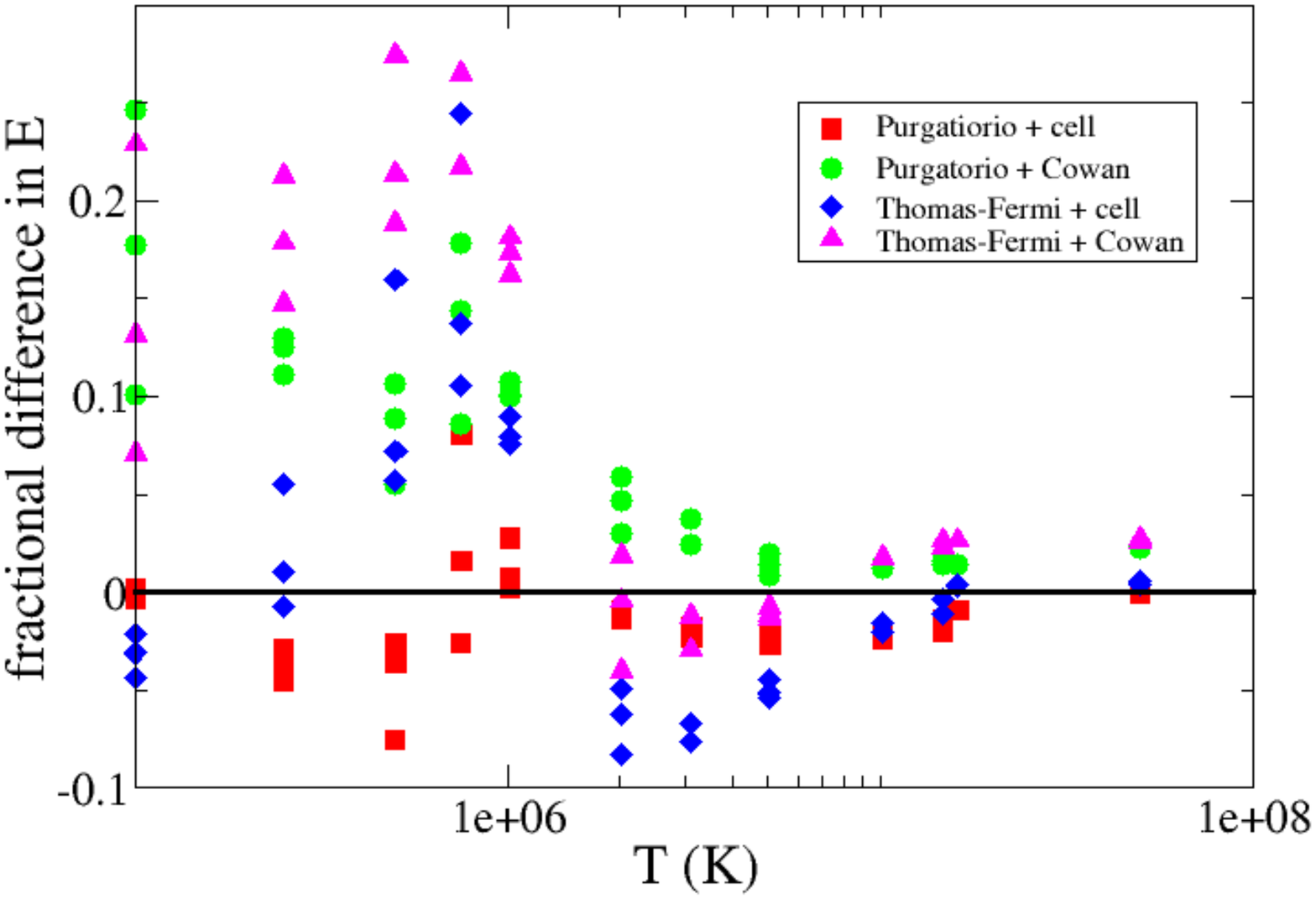}
\includegraphics[scale=0.40]{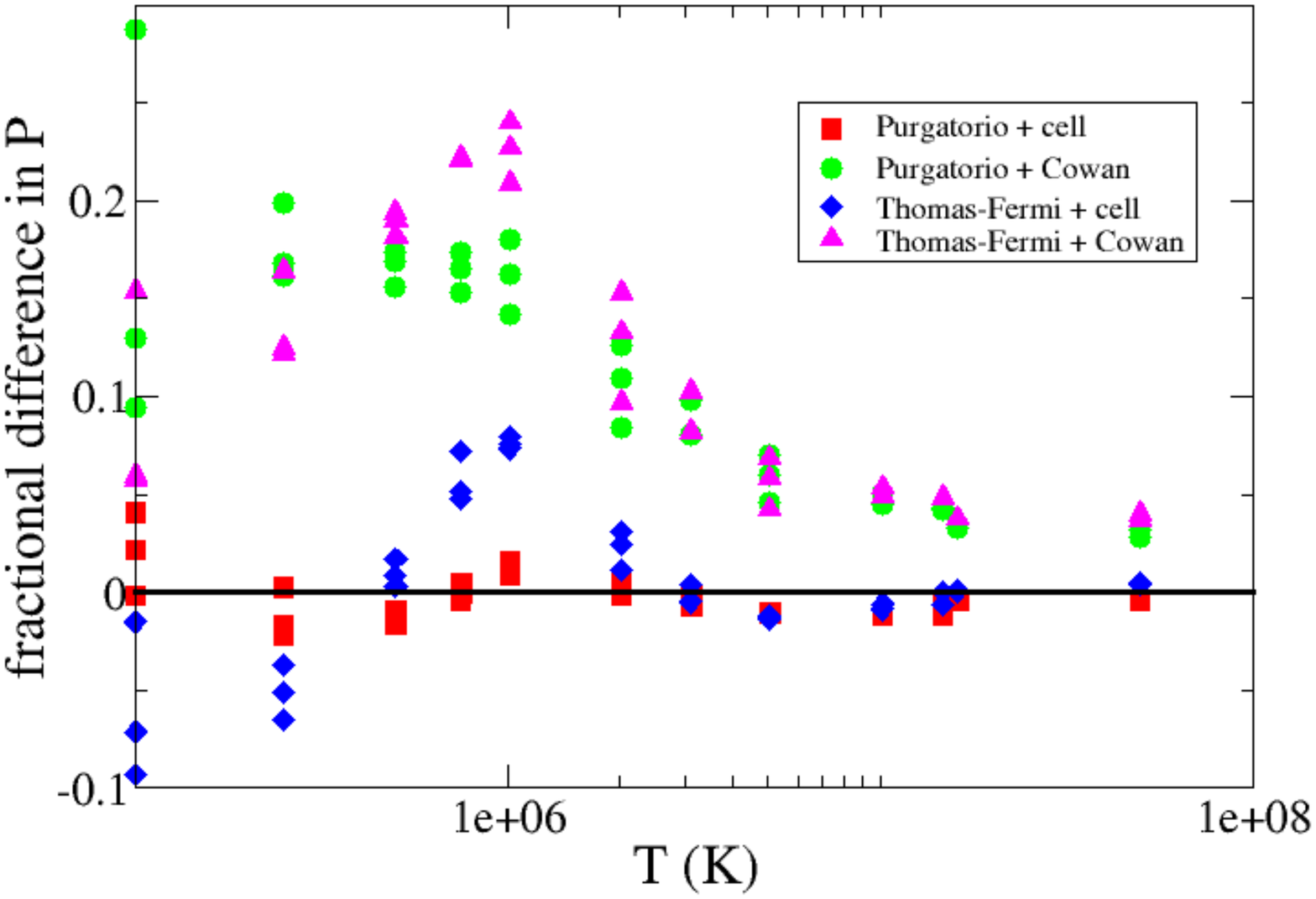}
\caption{Fractional differences in internal energy (a) and pressure (b) between PIMC results for the high-$T$ liquid/plasma and the results of our EOS model, assuming various combinations of ion-thermal and electron-thermal models. the x-axis is $T$ in K. For each liquid EOS model (indicated by color), three isochores are represented: $\rho=$ 3.18, 8.5, and 11.18 g/cc. See text for details.
}
\label{PIMC}
\end{figure}

To explore both possibilities while investigating a far wider range of $T$, we compare to our PIMC results for $E$ and $P$, as generated by the prescription outlined in Section II B. Fig. \ref{PIMC} shows the fractional differences, $(X_{\rm model} - X_{\rm PIMC})/X_{\rm PIMC}$ \cite{caveat_frac}, between liquid C EOS model predictions and the results of PIMC for three isochores, $\rho=$ 3.18 g/cc, 8.5 g/cc, and 11.18 g/cc. Both internal energy and pressure differences are displayed. For the liquid C EOS models, we consider four variants: 1. $F_{\rm ion}$= Cowan, $F_{\rm elec}= $ PURGATORIO (green), 2. $F_{\rm ion}$= Cowan, $F_{\rm elec}= $ Thomas-Fermi (magenta), 3. $F_{\rm ion}$= cell model, $F_{\rm elec}= $ PURGATORIO (red), 4. $F_{\rm ion}$= cell  model, and $F_{\rm elec}= $ Thomas-Fermi (blue). Note the exceptionally wide range of $T$ displayed in log-scale on the x-axis ($5\times 10^{5}$ K $\to 10^{8}$ K). Each set of points (of a given color/symbol) includes data from three isochores with densities: $\rho=$ 3.18 g/cc, 8.5 g/cc, and 11.18 g/cc. All model variants give results which are coincident with PIMC at sufficiently high-$T$. However, there are significant deviations in $E$ and $P$ for lower temperatures. Clearly, it is the red set of points, corresponding to $F_{\rm ion}=$ cell model, $F_{\rm elec}=$ PURGATORIO, which presents the smallest deviations from PIMC. Furthermore, while the preference for PURGATORIO over Thomas-Fermi when fixing the ion-thermal model is notable but modest, the preference for the cell model over the Cowan model is quite dramatic for both choices of $F_{\rm elec}$. Finally, it is interesting that there is a pronounced peak in these fractional differences at $T \sim 10^{6}$ K for both $E$ and $P$, and for all three isochores presented. It is possible that this arises from an ionization feature in the PIMC which is somehow manifested rather differently in the Thomas-Fermi and PURGATORIO models. However, it is also curious that these deviations attain their maxima at roughly the same $T$ even for very different densities, since one might expect such ionization features to be rather dependent on density. The detailed shape of these curves awaits further analysis. 

The conclusion of these comparisons is that both our DFT-MD {\it and} our PIMC results indicate that the rate of decay of $C_{V}^{\rm ion}$ to its ideal gas value is far too slow in the Cowan model, while the faster decay of the cell model provides far better agreement for liquid C in this range of densities (3 g/cc $< \rho <$ 12 g/cc). We therefore use the cell model together with the PURGATORIO electron-thermal model to build our liquid C free energy. 

\subsubsection{Final liquid free energy model}
\begin{figure}
\includegraphics[scale=0.40]{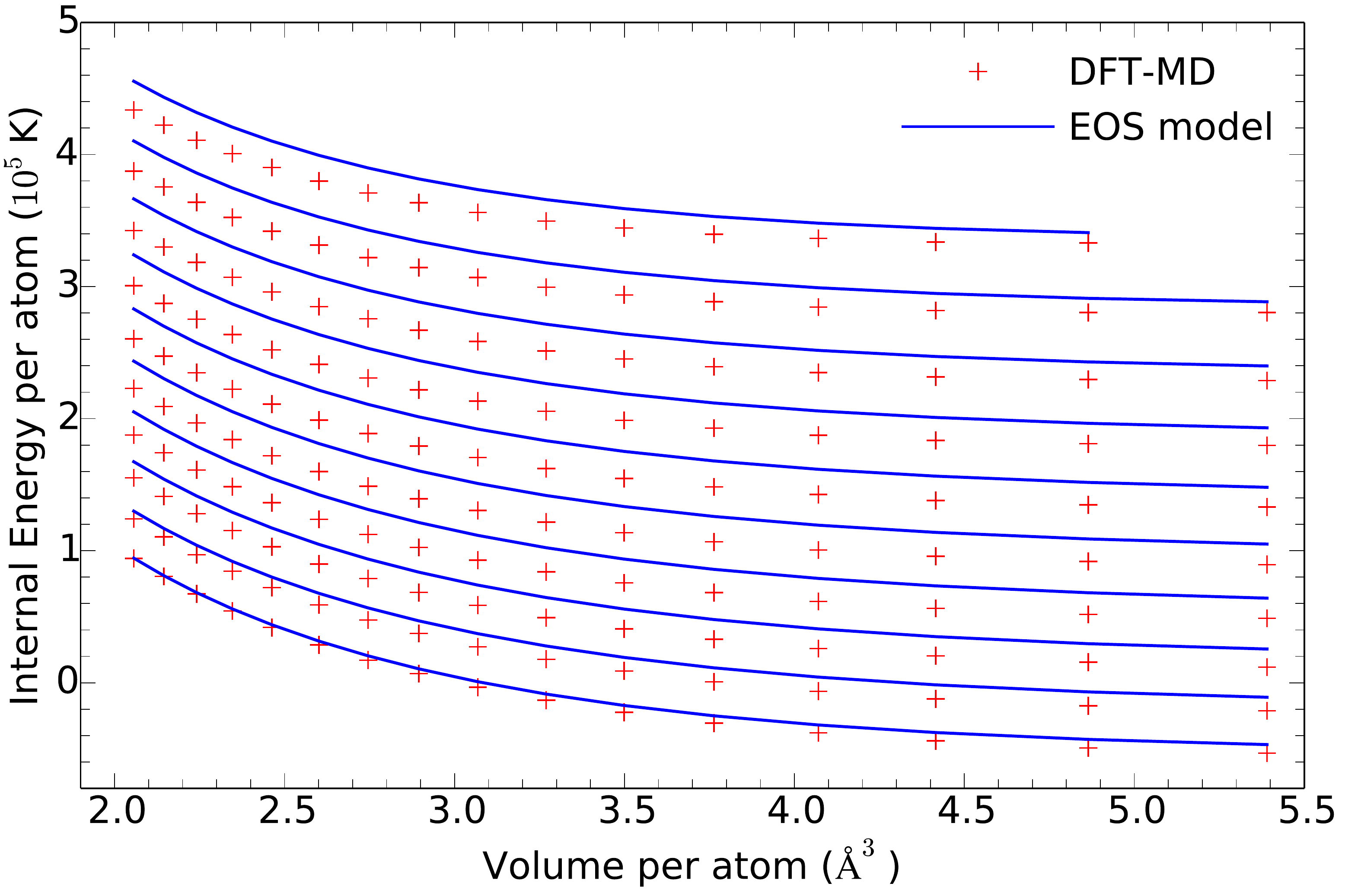}
\caption{Internal energy (in K/atom) isotherms for liquid C as computed by DFT-MD (red points) and our EOS model (blue lines). Temperatures range from 10,000 K to 100,000 K.}
\label{liqE}
\end{figure}

\begin{figure}
\includegraphics[scale=0.40]{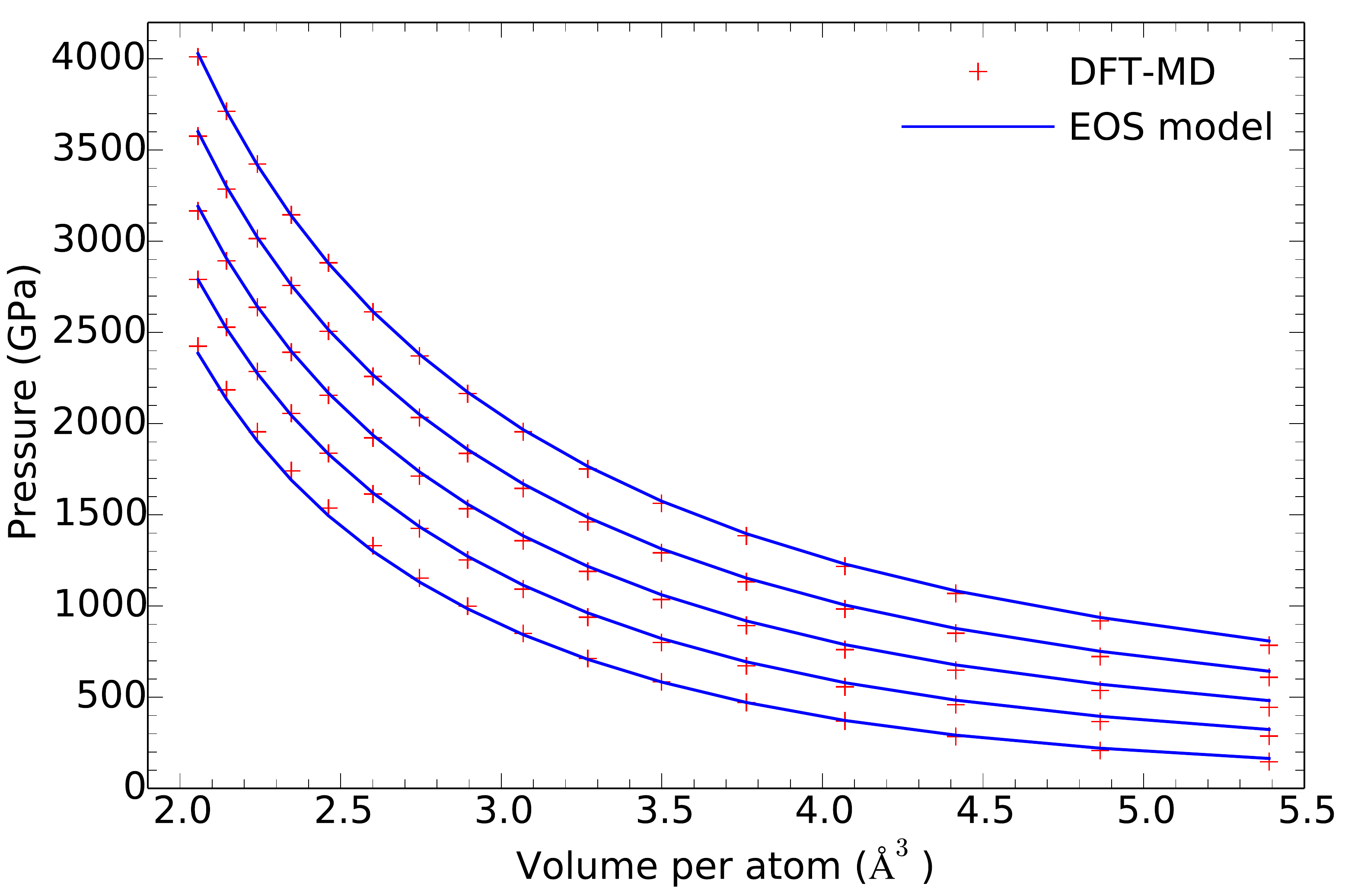}
\includegraphics[scale=0.40]{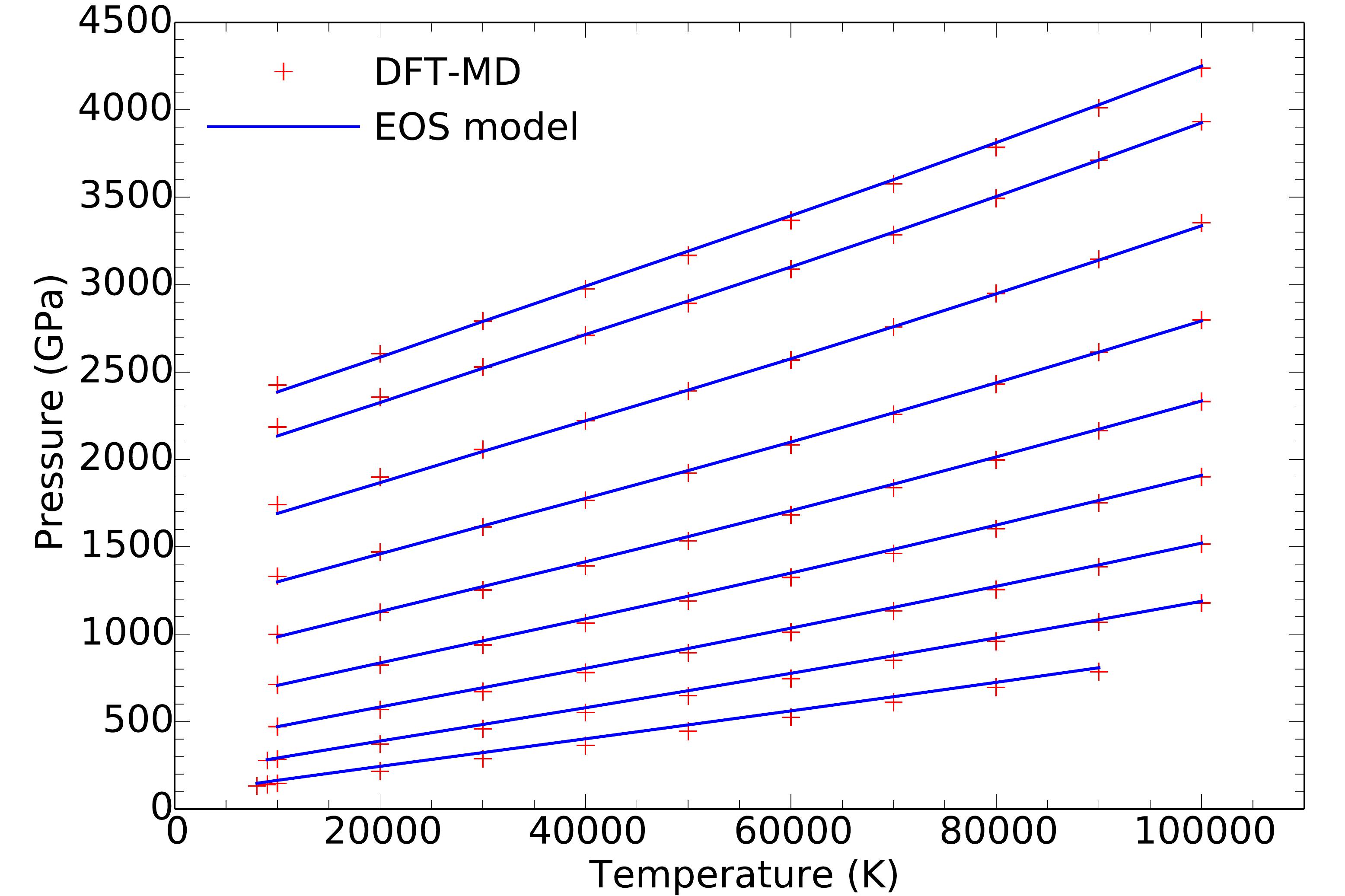}
\caption{(a) Pressure (in GPa) isotherms for liquid C as computed by DFT-MD (red points) and our EOS model (blue lines). (b) Pressure (in GPa) isochores for liquid C as computed by DFT-MD (red points) and our EOS model (blue lines). The range of $T$ represented in (a) can be seen from (b), and likewise for the range of $V$ in (b). Note that this plot contains roughly one-half of the isotherms and isochores in this density and temperature range for which we produced DFT-MD data.
}
\label{liqP}
\end{figure}

\begin{figure}
\includegraphics[scale=0.40]{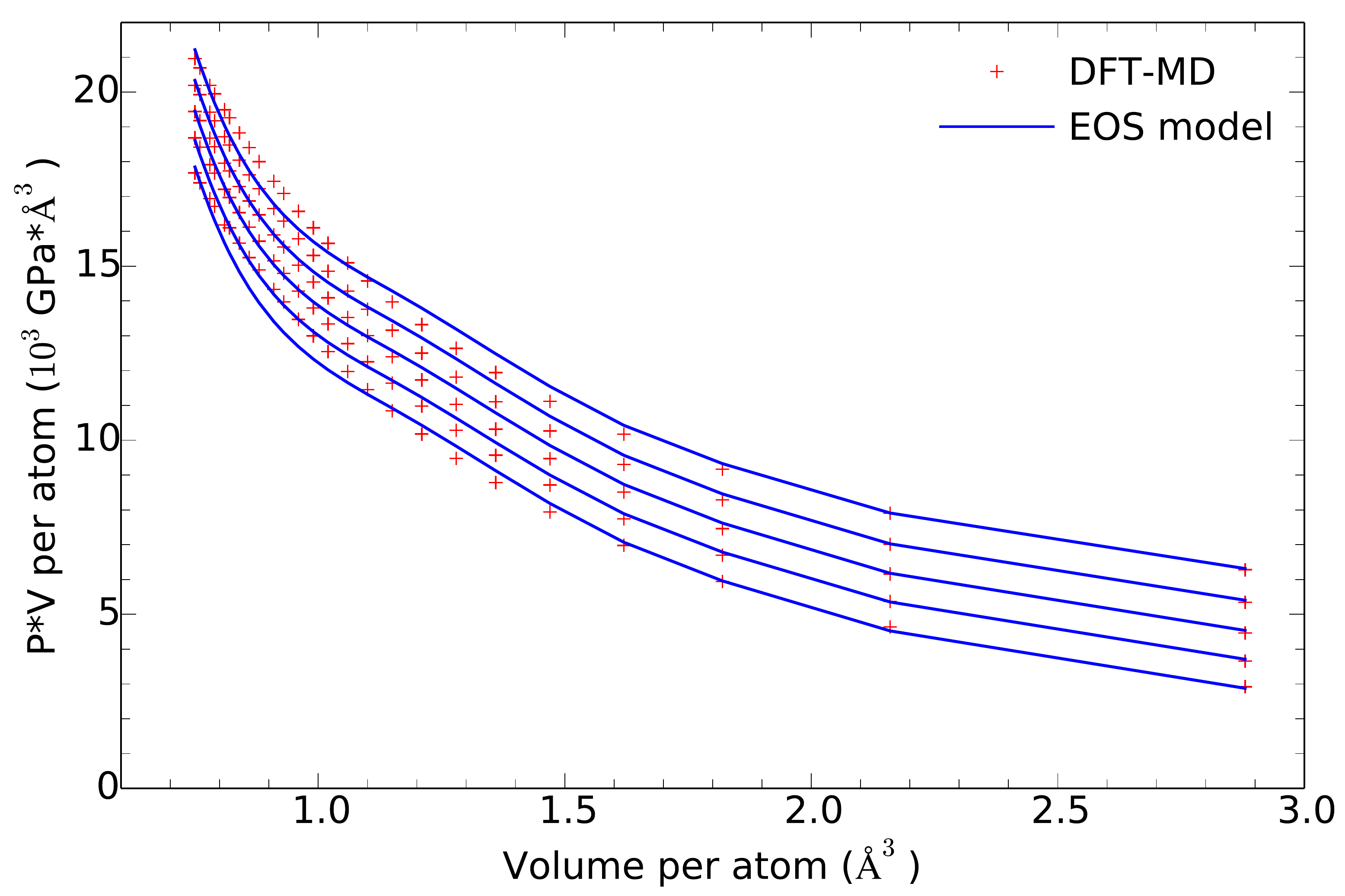}
\caption{Pressure times volume per atom (in GPa$\times \AA^{3}$) isochores for liquid C as computed by DFT-MD (red points) and our EOS model (blue lines). The range of $T$ represented is 10,000 K to 100,000 K. We plot $P\times V$ rather than simply $P$ here to show more clearly the data/model discrepancies at low-$V$. Note that this plot contains roughly one-half of the isotherms in this density range for which we produced DFT-MD data.}
\label{highPliqP}
\end{figure}

The parameters of our liquid cold curve (including break-points) and ion-thermal characteristic temperature are listed in Table 2. Minimal changes have been made to the liquid free energy model of Ref.\cite{Correa}. These changes include: 1. An altered cold curve at high compressions, 2. The use of the Purgatorio electron-thermal term, and 3. The use of the cell model addition to the Chisolm-Wallace Mie-Gr\"uneisen term (see above). The lower pressure ($< 25$ GPa) cold curve and the Debye-like ion-thermal model itself is unchanged from that of Ref.\cite{Correa}. Fig. \ref{liqE} shows isotherms of internal energy for the liquid as computed by both DFT-MD and our liquid EOS model. The range of $V$ is chosen to correspond to that for which diamond and BC8 phases are stable at lower temperatures. Note that while the lowest temperature ($T$= 10000 K) isotherm of the model and the DFT-MD are in reasonable accord, especially for smaller $V$, higher-$T$ isotherms deviate in a manner which depends on $V$. This results from the Chisolm-Wallace + cell + Purgatorio models being less than perfect for C, a fact which is not surprising given their relative simplicity. Still, as we have discussed, the agreement is far better than that obtained by using the Cowan and/or Thomas-Fermi free energy contributions (see Figs.\ref{ESH} and \ref{PIMC}). Pressure isotherms and isochores (Fig. \ref{liqP}) on the same grid of $(V,T)$ exhibit somewhat better overall agreement. The ultra-high compression pressure isotherms of the liquid are shown in Fig. \ref{highPliqP}; agreement between our model and our DFT-MD results is quite good but not perfect.

\begin{table}
\begin{center}
\begin{tabular}{| c | c |}
\hline
parameter & liquid \\ \hline
$V_{0}$ & 8.596 \\ \hline
$B_{0}$ & 51.11 \\ \hline
$B'$ & 5.848  \\ \hline
$\phi_{0}$ & -7.5 \\ \hline
$V_{p}$ & 6.695 \\ \hline
$\theta^{0}$ & 520.0 \\ \hline
$A$ & 0.0 \\ \hline
$B$ & 0.84 \\ \hline\hline\hline
$V_{b1}$ & 3.9 \\ \hline
$n_{1}$ & 3.0 \\ \hline
$a_{1}$ & -5.0 \\ \hline
$b_{1}$ & 5.0 \\ \hline
$V_{b2}$ & 2.7 \\ \hline
$n_{2}$ & 3.0 \\ \hline
$a_{2}$ & 10.0 \\ \hline
$b_{2}$ & 3.0 \\ \hline
$V_{b3}$ & 1.9 \\ \hline
$n_{3}$ & 2.0 \\ \hline
$a_{3}$ & -40.0 \\ \hline
$b_{3}$ & 5.0 \\ \hline
$V_{b4}$ & 1.13 \\ \hline
$n_{4}$ & 3.0 \\ \hline
$a_{4}$ & 80.0 \\ \hline
$b_{4}$ & 5.0 \\ \hline
\hline
\end{tabular}
\caption{EOS model parameters for the liquid phase of our multiphase C EOS. The upper segment of the Table concerns cold curve and ion-thermal parameters, and the lower segment contains break-points parameters used to further define the cold curve. All volumes ($V$) are in $\AA^{3}$/atom, $B_{0}$ is in GPa, $B'$ is unitless, $\phi_{0}$ is in eV/atom, all characteristic temperatures ($\theta$) are in Kelvins, $a$-parameters are in eV/atom, and $b$- and $n$-parameters are unitless. See Ref.\cite{bp} for the specification of the break-point formula.}
\end{center}
\end{table}

\subsection{phase diagram}
\begin{figure}
\includegraphics[scale=0.40]{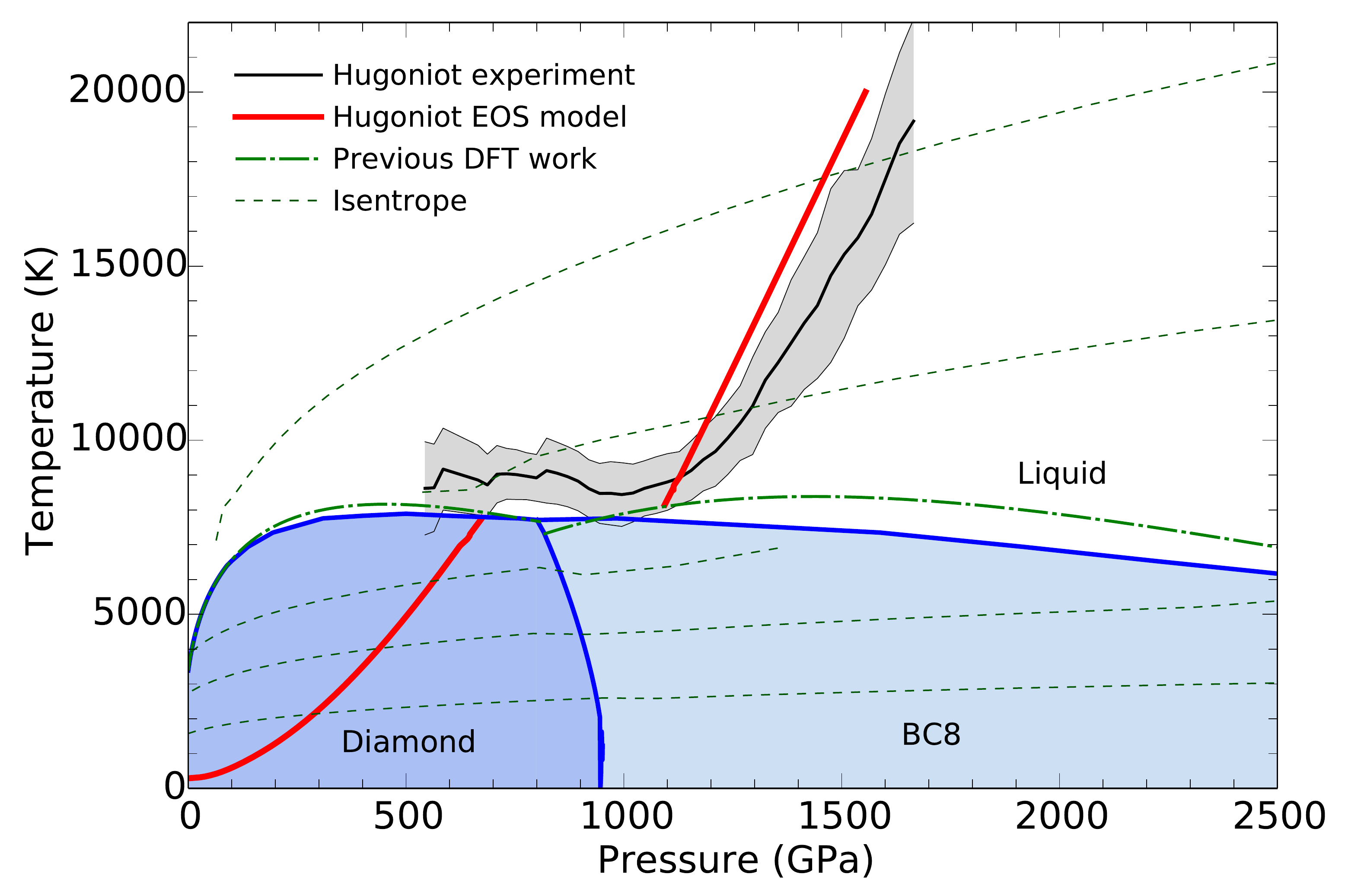}
\includegraphics[scale=0.40]{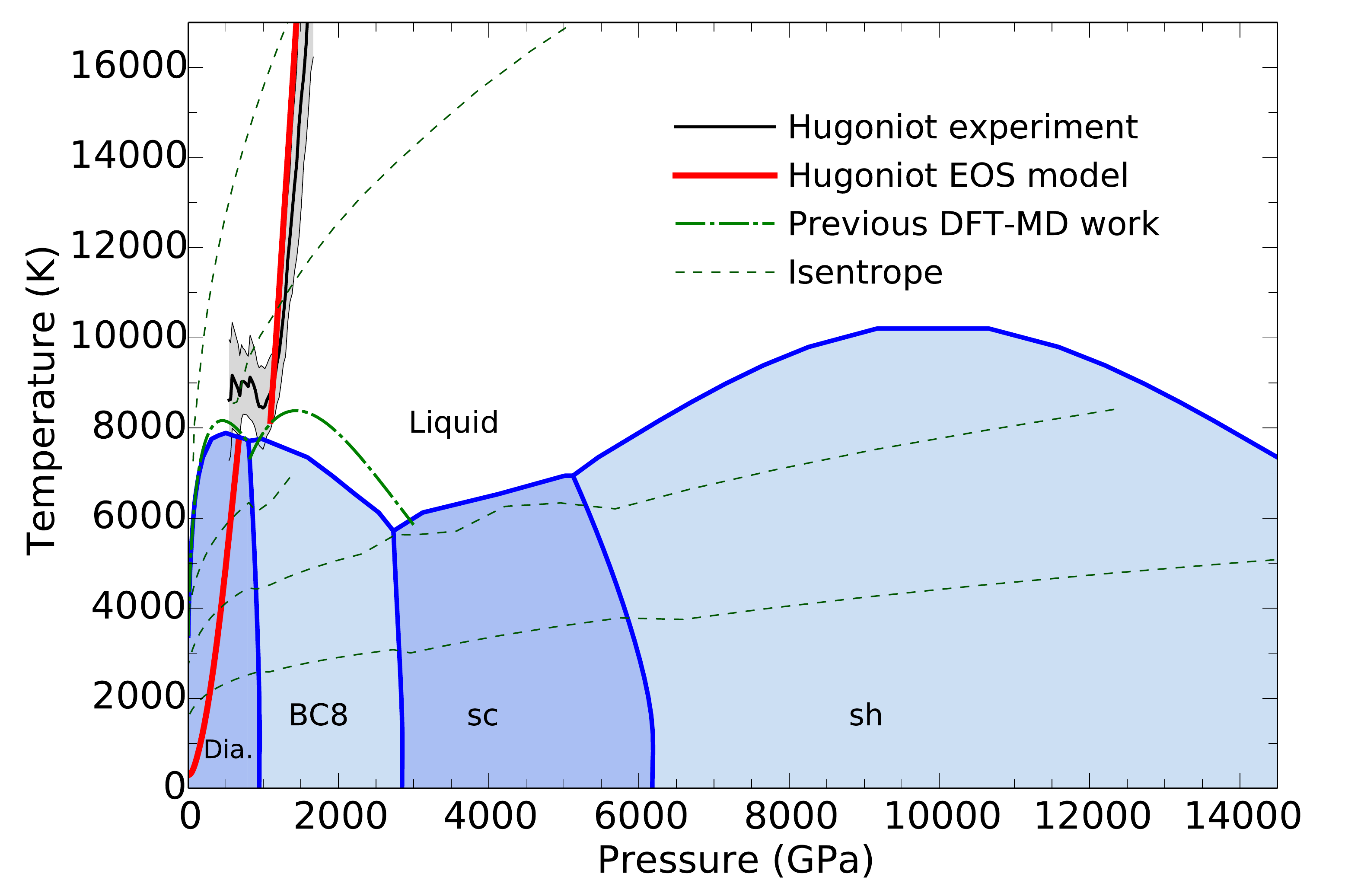}
\caption{Phase diagram of C within our EOS model (phase lines are indicated by thick blue lines) for narrower (a) and wider (b) ranges of pressure. Green dashed-dotted thin lines indicate Kechin fits to the diamond and BC8 melt curves as computed in the work of Ref.\cite{PNAS}. The red curve shows the principal Hugoniot as predicted by our model. The shaded grey region indicates the error bars for the principal Hugoniot of Ref.\cite{Eggert}. Predicted isentropes are indicated by dashed green lines; entropy values, increasing from the bottom of each figure, are: 3.78 $k_{\rm B}$/atom, 5.49 $k_{\rm B}$/atom, 6.61 $k_{\rm B}$/atom, 12.56 $k_{\rm B}$/atom, and 12.72 $k_{\rm B}$/atom.
}
\label{pd}
\end{figure}

Given the free energy models for the individual solid (diamond, BC8, sc, sh) and liquid phases, we construct the phase diagram and multiphase EOS by invoking the Maxwell construction for each pair of phases (here denoted 1 and 2), and at each temperature, $T$, of interest \cite{Wallace}:
\begin{equation}
F_{2}(V_{2},T) - F_{1}(V_{1},T)= -P(V_{2} - V_{1}),
\end{equation}
where $F_{1,2}$ are total free energies for each phase, $V_{1,2}$ are the transition volumes, and $P$ is the transition pressure. Since the individual solid phases are only meaningful (as in metastable) in the neighborhood of their fields of thermodynamic stability, we only consider each solid phase free energy function throughout a restricted range of $V$, so as to minimize the spurious effects of improper extrapolation. In particular, we consider diamond from 2.5 $\AA^{3}$/atom $\to$ 7.0 $\AA^{3}$/atom, BC8 from 1.3 $\AA^{3}$/atom $\to$ 3.5 $\AA^{3}$/atom, sc from 1.0 $\AA^{3}$/atom $\to$ 7.0 $\AA^{3}$/atom, and sh from 0.01 $\AA^{3}$/atom (undoubtedly well beyond its actually range of stability) $\to$ 7.0 $\AA^{3}$/atom. The liquid is considered throughout the wide range: 0.01 $\AA^{3}$/atom $\to$ 7.0 $\AA^{3}$/atom, though the actual intended range of validity of the EOS model reported here is limited to the regime covered by our {\it ab initio} calculations (see Supplementary Material): 0.75 $\AA^{3}$/atom $<$ $V$ $< $ 5.4 $\AA^{3}$/atom (as we stated above, we choose not to require agreement with the 0.1 g/cc PIMC isochore in this work).
Fig. \ref{pd}a shows the resulting phase diagram in the regime of pressure and temperature addressed in the work of Ref.\cite{Correa}. Our phase lines computed in this work differ only slightly from those in that earlier study. In particular, the diamond-BC8 transition pressure is slightly lower; this is due to our diamond and BC8 cold curves being slightly different from those of Ref.\cite{Correa} (higher plane wave cutoffs and different pseudopotentials were used here). Also, the BC8-phase melt temperature is slightly lower, as evidenced by our comparison to the Kechin fits to the carbon melt curves of Ref.\cite{PNAS} to which the EOS model of Ref.\cite{Correa} was fit. Nevertheless, our phase diagram is quite close to that of the older model in the range of its applicability. Also shown in Fig. \ref{pd} are various isentropes as computed with our multiphase EOS model; values of the entropy for each are listed in the figure caption.

The principal Hugoniot of carbon computed with the model is shown as well in Fig. \ref{pd}. This is defined to be the locus of final states accessible via a planar one-dimensional shock, given a particular assumed initial state, here taken to be $\rho=$ 3.52 g/cc and $T=$ 300 K. This curve has two branches: one for which the final state is the diamond phase, and one in which the liquid is the final state. They are separated by a flat region in $T$ straddling the pressure at which the diamond $\to$ BC8 transition is predicted to occur. In this portion of the Hugoniot curve the final state is in a mixed-phase region, the size of which is directly related to the latent heat of melting, as discussed in Ref.\cite{Correa}. We predict a principal Hugoniot which is similar to that predicted by the earlier carbon EOS model \cite{Correa} (indeed, it is nearly identical in the diamond-phase portion). However, there is a pronounced increase ($\sim$ 100 GPa) in the Hugoniot final state pressure of the liquid portion, which is explained by the added latent heat of melting resulting from the addition of the PURGATORIO electron-thermal contribution. Our change in entropy from solid to liquid at constant $V$ at the shock melt conditions is $\sim 3.2 k_{\rm B}$/atom, while in the model of Ref.\cite{Correa} it is $\sim 2.9 k_{\rm B}$/atom. This in turn moves our liquid branch of the Hugoniot a bit closer to the measured $T$ vs. $P$ Hugoniot data of Eggert et al.\cite{Eggert} (the upper and lower bounds of their experimental error bars are displayed as well in Fig. \ref{pd}a). Our Hugoniot still fails to fall within the bounds of their measurements, however.

The phase diagram of our multiphase C EOS model over a larger range of pressure and temperature is shown in Fig. \ref{pd}b. The higher-$P$ solid phases, sc and sh, appear as well, in addition to diamond, BC8, and liquid phases. Several important features are worth noting: 1. The BC8 $\to$ sc transition pressure is nearly independent of temperature, as is the case for the diamond $\to$ BC8 transition pressure. This is because the relevant moments,  $\theta(V)$, of the PDOSs of the phases on either side of each transition are very similar at the densities at which the transitions occur, as discussed above. 2. The sc $\to$ sh phase line is less vertical, and curves to lower-$P$ for higher $T$, but we do {\it not } predict that this phase line intersects the BC8-sc phase line before intersecting the melt curve. This is in stark contrast to the prediction of Ref.\cite{MC}, which shows a sc-sh phase line which intersects the BC8-sc line at $T \sim 5000$ K. Though we use similar ab initio and EOS modeling methods to theirs, we suspect that their treatment of the $T$-dependence of the transition pressure in their Gibbs free energy matching prescription is sufficiently different from our approach to produce this discrepancy. We have checked that alternate ways of fitting our EOS data (cold curves, PDOSs and resulting ion-thermal free energy terms) still give the general phase diagram topology we present here, though we do not discount the possibility that the extreme sensitivity of the phase lines to small changes in the individual phase free energies can render our predictions somewhat inaccurate, especially at these higher pressures where the $F(V)$ curves are nearly parallel. 3. The value of the sc-sh transition pressure at low-$T$ is $\sim 20 \%$ higher than that reported in Ref.\cite{MC}, though the position of the BC8-sc transition line is in reasonable agreement between the two studies. This is likely a result of the use of different pseudopotentials and plane wave convergence criteria. Again, the nearly coincident free energy functions for different phases at high pressures makes the phase lines extremely sensitive to any otherwise subtle changes. 4. The melt line of the sh phase possesses a maximum at $P \sim 10,000$ GPa. This is striking, but is one of the least certain predictions we make in this work; we have seen that slight changes to our EOS model parameters which produce small changes to our liquid free energy (which are within both our assumed model and DFT-MD uncertainties) yield sc melt temperatures which exhibit markedly different behavior. 

\begin{figure}
\includegraphics[scale=0.50]{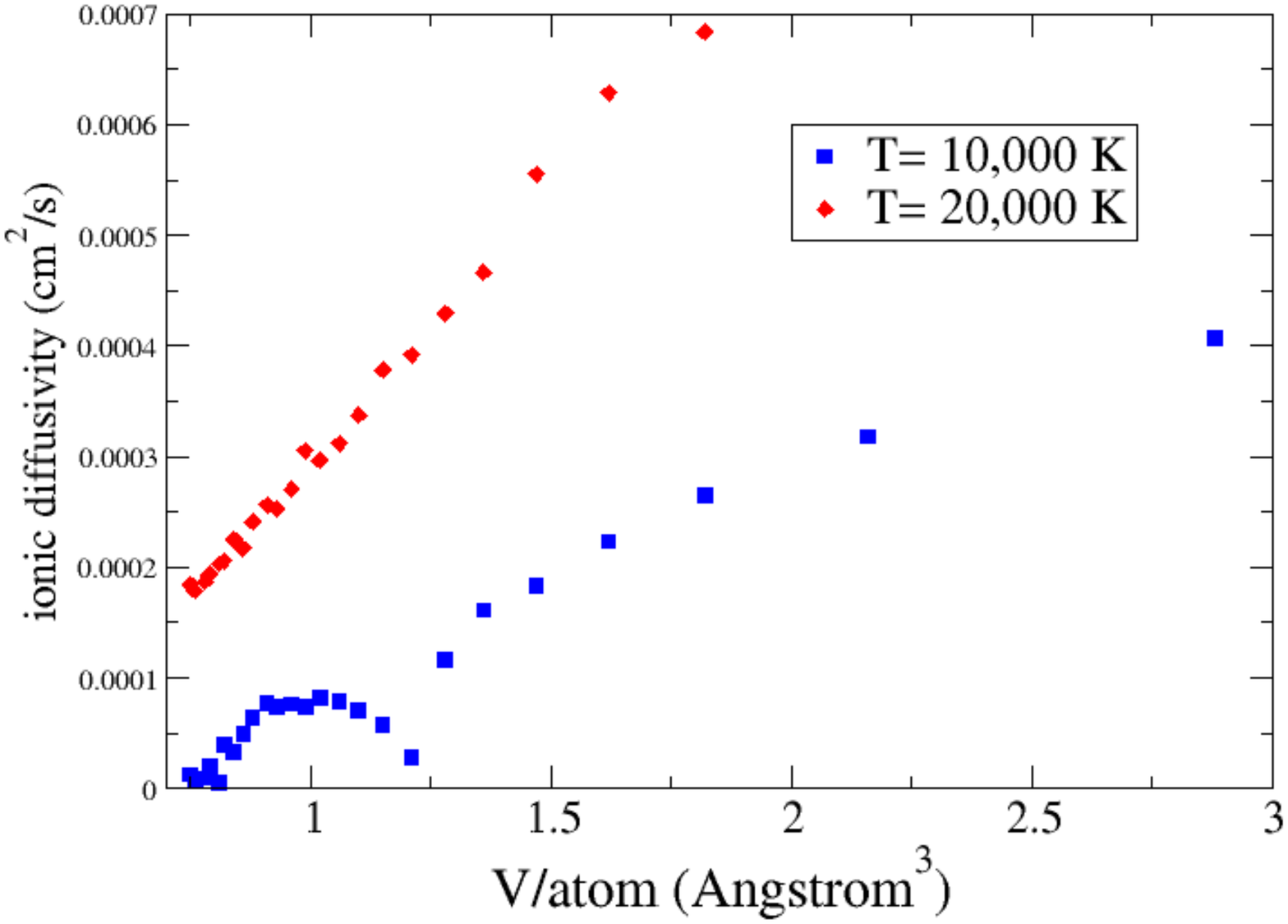}
\caption{Ionic diffusivity for liquid C in cm$^{2}$/s vs. $V$ in $\AA^{3}$/atom for two isotherms: $T=$ 10,000 K (lower) and $T=$ 20,000 K (upper), as computed by DFT-MD.}
\label{diff}
\end{figure}

Our reason for favoring this particular EOS model parameterization over other nearly equivalent ones can be found in our computation of the ionic diffusivities using the DFT-MD. Fig. \ref{diff} shows two diffusivity isotherms, the lower one for $T=$ 10,000 K and the upper one for $T=$ 20,000 K. A very small (approaching zero) diffusivity is indicative of a solid phase. The ionic diffusivity at 20,000 K shows the expected decrease as $V$ decreases, but is otherwise large as expected for a liquid. The diffusivity at 10,000 K shows a pronounced dip towards zero at a $V \sim$ 1.2 $\AA^{3}$/atom, corresponding to a pressure of roughly 10,000 GPa (see Fig. \ref{highPliqP}). This is a clear indication of solidification, and indeed, the phase diagram of Fig. \ref{pd}b shows a melt curve which attains its maximum (of just over 10,000 K) at $P= 10,000$ GPa. Though this provides a nice consistency check on our DFT-MD-derived C EOS model, we caution that the diffusivity calculations were performed in a cell of 64 atoms (as were the DFT-MD calculations of EOS), and solidification can be biased in simulations with small cells of fixed shape \cite{caveat_melt}. 

The overarching points to be made about our prediction of the phase diagram of C are two-fold: First, our predictions of phase lines are necessarily less accurate the higher the pressure is, for the reasons mentioned directly above. These inaccuracies are the combined result of the inherent uncertainties in the underlying ab initio electronic structure data, and the deficiencies of EOS models which have a finite number of adjustable parameters. Second, the predicted volume changes across such high-$P$ phase lines are often exceedingly small (which is what makes the accurate determination of these transition lines difficult). Thus, the least accurate phase lines have a minimal impact on the EOS itself ($P$ and $E$ at a given $\rho$ and $T$).

\section{Comparisons to experimental results}
We have already discussed the comparison of the results of our C EOS model to principal Hugoniot data in the $(P,T)$-plane. These recent measurements made use of optical pyrometry techniques \cite{Eggert} which lie outside the typical purview of classic shock-compression experiments. Now we compare to a larger set of principal Hugoniot data for elemental carbon transformed more typically to the $(\rho,P)$-plane, as well as to room-temperature isotherm data from static compression experiments.

\begin{figure}
\includegraphics[scale=0.50]{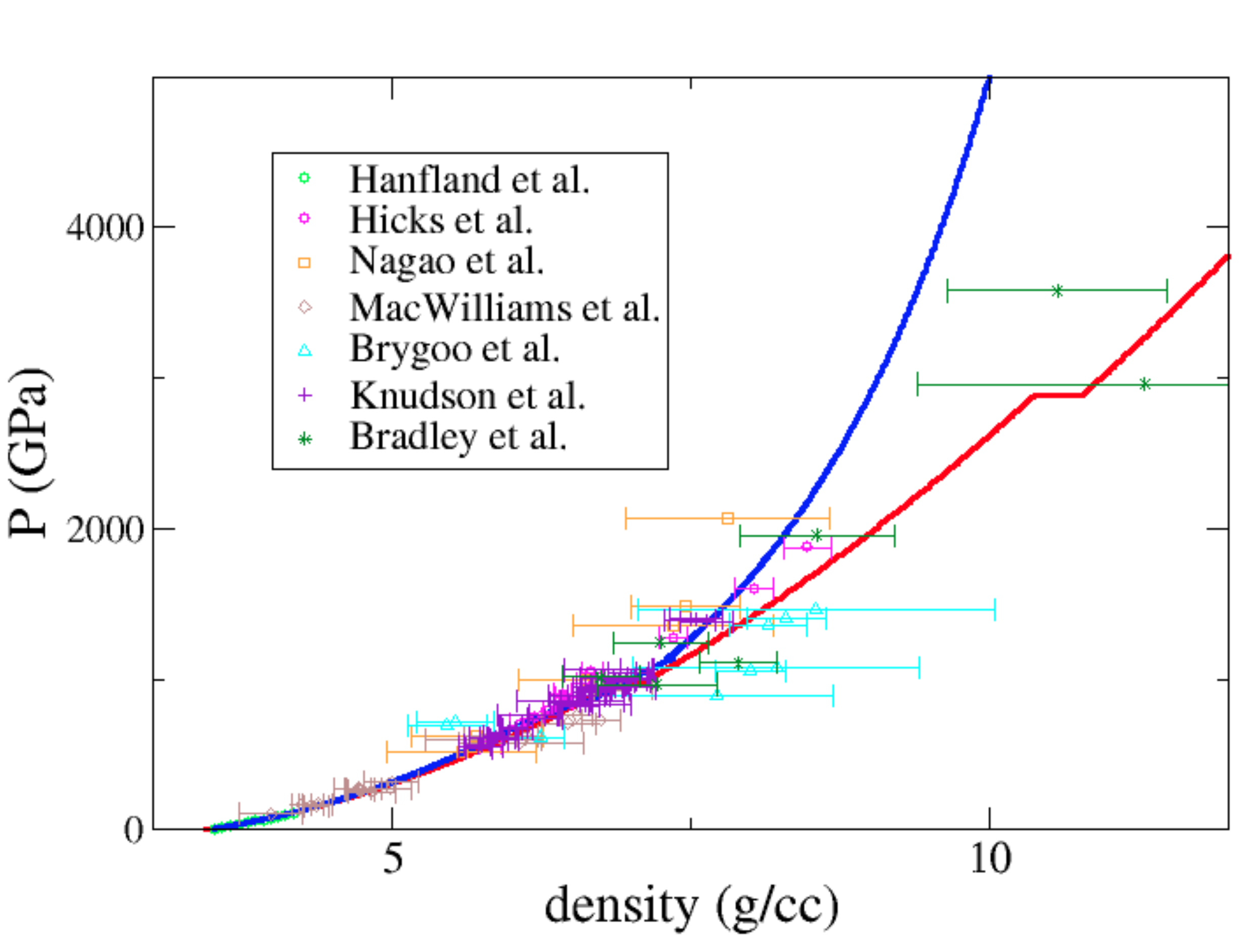}
\caption{300 K isotherm (red curve) and principal Hugoniot (blue curve) as computed by our EOS model. Green points indicate static compression data from Ref.\cite{isotherm}, and all other symbols (+ error bars) indicate shock data from Refs.\cite{Bradley,Nagao,Hicks,Brygoo,McWilliams,Z}.}
\label{RH}
\end{figure}

Fig. \ref{RH} shows the 300 K isotherm (red) and the principal Hugoniot (blue) in $(\rho,P)$-space as calculated from our multiphase C EOS model. The isotherm (which is very similar to the principal isentrope) shows flat regions where the phase transitions occur (BC8 $\to$ sc and sc $\to$ sh in this plot). The $P$ vs. $\rho$ Hugoniot is nearly indistinguishable from the isotherm below roughly 1000 GPa, but deviates dramatically from it at higher stresses. The point at which they diverge corresponds to the point where the Hugoniot final states begin to reside in the liquid phase (see Fig. \ref{pd}). Also shown on this figure are data from static compression measurements of the room-temperature isotherm \cite{isotherm} (green circles at the lowest $P$), magnetically-driven flyer plate studies \cite{Z} (dense set of magenta + symbols with $\rho \sim$ 6 - 7 g/cc) and multiple sets of laser-shock data on the principal Hugoniot of diamond (symbols with error bars) \cite{Bradley,Nagao,Hicks,Brygoo,McWilliams}. Note that the much of the highest-$P$ Hugoniot data \cite{Bradley} seems to straddle the 300 K isotherm, even well above $P= 1000$ GPa; indeed, the data of Brygoo et al. \cite{Brygoo} even falls below our prediction of the room-$T$ isotherm. This is very puzzling in light of our theoretical results which suggest that the Hugoniot should be much stiffer than the isotherm at these larger compressions. Our previous ab initio-based C EOS \cite{Correa} was only fit to DFT-MD data up to $T=$ 20,000 K, which translates to $P=$ 1600 GPa on the principal Hugoniot. At this pressure, the Hugoniot and the 300 K isotherm are still rather close. The EOS model of this work, however, is validated by comparing to calculations (DFT, PIMC) that span the full range of temperature up to ideal gas conditions; note in particular the comparison to DFT-MD data on a dense grid of $(\rho,T)$ as pictured in Figs. \ref{liqE} and \ref{liqP} which covers the entire range of conditions relevant for the comparison of Fig. \ref{RH}. 

In order to check the robustness of our conclusions that the principal Hugoniot should indeed be far stiffer than the 300 K isotherm for $P >$ 1000 GPa, we have verified that the different variants of liquid C EOS we discussed in Section III.B.1 (Cowan + Purgatorio, cell + Thomas-Fermi, etc.) all show this same basic relationship between the low-$T$ isotherm and the Hugoniot. Looking back again to the experimental results pictured in Fig. \ref{RH}, we note that there are data with $P >$ 1000 GPa that seem to be closer to our Hugoniot predictions, in particular those of Nagao et al. \cite{Nagao} and Hicks et al. \cite{Hicks}. It must be mentioned that the Hicks et al. data shown here is {\it not} that as presented in the original Hicks et al. reference, but is instead a data set which results from a reanalysis \cite{Smith} of the quartz EOS \cite{quartz} which was used as an impedance-matched standard in that work. The original uncorrected data set \cite{Hicks} lies much closer to our low-$T$ isotherm than to our principal Hugoniot prediction in this higher-$P$ region. These experimental improvements notwithstanding, we still submit that further work must be done to resolve the theory-experiment discrepancy at the very highest compressions shown here.

\section{Conclusions}
We have constructed a 5-phase EOS for elemental carbon based entirely on ab initio electronic structure calculations of the density functional theory and path integral Monte Carlo varieties. The PIMC and high-temperature DFT-MD data were particularly useful in helping us differentiate between different choices of ion-thermal (Cowan, cell) and electron-thermal (Thomas-Fermi, PURGATORIO) models. We found that these data strongly favor the cell + PURGATORIO combination. The multiphase EOS is constructed entirely without patching different models together in an ad hoc fashion; rather, full thermodynamic consistency is maintained and the appropriate high-$T$ limit is reached in a seamless manner. 

Our EOS model includes, in addition to diamond, BC8, and liquid, two of the ultra-high pressure solid phases recently predicted by Martinez-Canales et al. \cite{MC}: simple-cubic, and simple-hexagonal. Though our prediction of the phase diagram of carbon in this high-$P$ region is broadly similar to theirs, the detailed positions of phase lines and resulting triple-points are somewhat different, owing to the sensitivity of the phase lines to the detailed prescriptions for obtaining the individual phase free energies and computing the transition pressures. Further work should be done to resolve these details, though we maintain that the EOS itself should be reasonably accurate at high compressions in spite of this.

Comparison to recent laser-shock compression data \cite{Bradley,Nagao,Hicks,Brygoo,McWilliams} shows notable disagreement with the subset of results \cite{Bradley,Brygoo} which suggest that the principal Hugoniot is nearly coincident with the low-$T$ isotherm even at pressures in excess of 1000 GPa. It is not at all clear at the moment as to what is causing such a discrepancy, particularly since the general features of our predictions (seen in Fig. \ref{RH}) are quite robust and seemingly independent of many of the choices we have made in modeling the EOS of liquid C. We are encouraged, however, that a more recent reanalysis \cite{Smith} of the data of Hicks et al. \cite{Hicks} seems to bring their principal Hugoniot into better agreement with our prediction.

One of the central results of this work is that our high-$T$ DFT-MD and PIMC data on carbon are found to be in excellent agreement with a new ion-thermal free energy model. This so-called cell model exhibits an evolution from Dulong-Petit to ideal gas limits as $T$ is increased which is more rapid than that of the older Cowan model. Although the results presented here are specific to carbon, existing EOS models for other materials that are based on the widely-used Cowan model may need to be re-examined in this light, as those materials may also possess a somewehat rapid decay of $C^{\rm ion}_{V}$ with $T$. Recent work with simpler classical inter-ionic potentials suggests that such rapid decays may be common \cite{LJ}.

\section{Acknowledgements}
We thank C.J. Pickard, M. Martinez-Canales, J. DuBois, M.A. Morales, G.W. Collins, P.A. Sterne, H.D. Whitley, D.M. Sanchez, J.I. Castor, D. Ho, D. Braun, and R.F. Smith for helpful discussions. 
{\small This work was performed under the auspices of the U.S. Department of Energy by Lawrence Livermore National Laboratory under Contract No. DE-AC52-07NA27344.}

\newpage



\begin{thebibliography}{100}
\bibitem{lowrho}The PIMC technique works efficiently even at quite low densities because no plane wave or similar expansion is needed, as employed in most DFT-MD codes.
\bibitem{caveatelec}It is important to note that GGA-DFT generally underestimates the electronic excitation gaps of materials, resulting in metalization pressures that are too low, and electronic excitation free energies (for systems on the border between insulating and metallic) which are too large. In this sense, our predictions of the effects of electronic excitation on the EOS of solid C are likely to be overestimates.
\bibitem{caveat_anh}For all solid phases whose models possess a $T^{2}$ (electron-thermal or anharmonic ion-thermal) free energy term, we apply corrections above $T$= 20,000 K, well beyond their ranges of stability, which smoothly force $F_{\rm ion}$ to a constant independent of $T$. This is done to prevent these solid phases from becoming more stable than the liquid at high-$T$, which would eventually occur if the $T^{2}$ term, meant strictly as a low-$T$ correction, is allowed to persist at high temperatures.
\bibitem{caveatTm}Note that $T_{\rm M}(V)$ is {\it not} the melting temperature of C (as in the original Cowan prescription used in Ref.\cite{QEOS}), but is rather the temperature above which the ion-thermal specific heat begins to deviate from the Dulong-Petit limit of $3k_{\rm B}$ and head toward the ideal gas value of $3/2 k_{\rm B}$ \cite{MGextension}. Note also that the conclusions we draw in this work regarding the lack of suitability of the Cowan model also apply if we lower the value of $T_{\rm M}(V_{0})$ to 10,000 K, which is just above the melt temperature of C over a wide range of compressions.
\bibitem{caveat_frac}Our definitions for the fractional differences in $E$ and $P$ would be problematic if $E_{\rm PIMC}$ or $P_{\rm PIMC}$ were close to zero for some $T$; this is not the case for the $T$ and $\rho$ considered here (and given the definition of the zero of internal energy employed).
\bibitem{caveat_melt}We have also computed the internal energy of an isochore with $V= 1.21 \AA^{3}$/atom using DFT-MD, and have shown that there is a pronounced jump in $E$ as $T$ is raised above 10,000 K, which again is right in the neighborhood of the maximum of the melt curve of the sh phase. For this set of calculations, a periodic cell with 392 C atoms was used.
\end{thebibliography}
\end{document}